\documentstyle[12pt,axodraw]{article}
\newcommand{\NP}{Nucl. Phys. }
\newcommand{\PR}{Phys. Rev. }
\newcommand{\PRL}{Phys. Rev. Lett. }
\newcommand{\PL}{Phys. Lett. }

\newcommand{\gmunu}{g_{\mu\nu}}
\newcommand{\Aalf}{A_{\alpha}}\newcommand{\Abet}{A_{\beta}}
\newcommand{\Agam}{A_{\gamma}}
\newcommand{\Zalf}{Z_{\alpha}}\newcommand{\Zbet}{Z_{\beta}}
\newcommand{\Zgam}{Z_{\gamma}}
\newcommand{\Amu}{A_{\mu}}\newcommand{\Anu}{A_{\nu}}
\newcommand{\Zmu}{Z_{\mu}}\newcommand{\Znu}{Z_{\nu}}
\newcommand{\Wpmu}{W^+_{\mu}}
\newcommand{\Wmnu}{W^-_{\nu}}

\newcommand{\deltae}{\Delta_{\epsilon}}

\begin{document}
\baselineskip=20pt

\pagenumbering{arabic}

\vspace{1.0cm}
\begin{flushright}
LU-ITP 2001/023
\end{flushright}

\begin{center}
{\Large\sf One loop renormalization of spontaneously broken
$U(2)$ gauge theory on noncommutative spacetime}\\[10pt]
\vspace{.5 cm}

{Yi Liao}
\vspace{1.0ex}

{\small Institut f\"ur Theoretische Physik, Universit\"at Leipzig,
\\
Augustusplatz 10/11, D-04109 Leipzig, Germany\\}

\vspace{2.0ex}

{\bf Abstract}
\end{center}

We examine the renormalizability problem of spontaneously broken
non-Abelian gauge theory on noncommutative spacetime. We show by
an explicit analysis of the $U(2)$ case that ultraviolet 
divergences can be removed at one loop level with the same limited
number of renormalization constants as required on commutative 
spacetime. We thus push forward the efforts towards constructing
realistic models of gauge interactions on noncommutative spacetime.

\begin{flushleft}


\end{flushleft}

\newpage
\begin{center}
{\bf 1. Introduction}
\end{center}

There have been intense activities in noncommutative (NC) field theory 
since it was found to arise naturally as a specific limit of string theory 
$\cite{string}$. But NC field theory is also interesting in its own
right both as a theory which may be relevant to the real world
and as a quantum structure on NC spacetime which is very distinct from the
one built on the ordinary commutative spacetime. Concerning this, we may 
mention two issues among others, the unitarity and causality problem 
$\cite{unitarity}$ and the ultraviolet-infrared mixing $\cite{mixing}$. 
The new interactions and Lorentz violation introduced by noncommutativity
also lead to novel phenomenological implications $\cite{pheno}$, some of 
which are quite different from those of ordinary new physics beyond the 
standard model.

Supposing the idea of NC spacetime is physically relevant, it would be
desirable to consider whether it is possible to extend the standard model 
of the electroweak and strong interactions to NC spacetime. It is not
completely clear at the moment how to construct such a realistic model due 
to restrictions on gauge groups from the closure of algebra 
$\cite{closure}$, on possible representations that can be well defined
$\cite{nogo}$ for a product of gauge groups, and potential obstacles in 
anomaly cancellation that becomes more restrictive $\cite{anomaly}$, 
and so on $\cite{susy}$. 
But it is of no doubt that such a model should be perturbatively 
renormalizable so that gauge symmetry can be maintained order by 
order in the renormalized theory. Although there is no general proof on
renormalizability of gauge theory on NC spacetime for the time being, 
explicit analyses are indeed available up to some order in some models.
It has been shown that exact $U(1)$ $\cite{u1}$ and $U(N)$ $\cite{un}$ 
gauge theories are renormalizable at one loop. The renormalizability of
the real $\phi^4$ theory has even been confirmed up to two loops 
$\cite{phi4}$. But theories with spontaneous symmetry breaking are more 
subtle. At first glance one might imagine that the divergence problem
cannot be worse compared to the commutative case because of oscillating
factors introduced by the star product. But actually renormalizability 
depends on delicate cancellation of seemingly different sources of
divergences which is governed by Ward identities. It is not self-evident
at all whether this well-weighted arrangement still persists in NC 
theories. This is especially true of spontaneously broken theories. 
As pointed out in Ref. $\cite{campbell}$, there are already 
problems for spontaneously broken global symmetries; namely, the 
Goldstone theorem holds valid at one loop level for the NC $U(N)$ 
linear $\sigma$ model with a properly ordered potential, but not for 
the $O(N)$ one (except for $N=2$ which is equivalent to $U(1)$). An 
explicit analysis for the spontaneously broken $U(1)$ gauge theory has 
been given recently in Ref. $\cite{petriello}$ with the positive 
result that the divergences can be consistently subtracted at one loop.
It is the purpose of this work to pursue further along this line by 
examining the non-Abelian case. The motivation for this should be 
clear from the above discussion; it is the non-Abelian case, 
especially $U(2)$, that is closer to our goal of building up a model 
of electroweak interactions on NC spacetime. Our positive result 
should be encouraging to the efforts in this direction.

The paper is organized as follows. In the next section we present the
setup of the model in which we will work. In particular we do 
gauge-fixing and work out its corresponding ghost terms including
counterterms. Section 3 contains the explicit result of divergences in
one loop 1PI functions. The renormalization constants are then determined
in the MS scheme. We summarize in the last section and state the 
limitations of this work and prospects for further study. The Feynman
rules are listed in Appendix A, and collected in Appendix B are 
topologically distinct Feynman diagrams for the 1PI functions computed 
in section 3.

\begin{center}
{\bf 2. The model}
\end{center}

{\it 2.1 Brief introduction to NC field theory}

Throughout this paper, by $n$ dimensional noncommutative spacetime we 
mean the one that satisfies the following canonical relation,
\begin{equation}
[\hat{x}_{\mu},\hat{x}_{\nu}]=i\theta_{\mu\nu},
\end{equation}
where $\theta_{\mu\nu}$ is a real, antisymmetric, $n\times n$ 
constant matrix. Following Weyl we can define a function on NC 
spacetime by the Fourier transform,
\begin{equation}
\hat{f}(\hat{x})=
\frac{1}{(2\pi)^{n/2}}\int d^4k~e^{ik_{\mu}\hat{x}^{\mu}}
\tilde{f}(k),
\end{equation}
where the same $\tilde{f}(k)$ simultaneously defines a function on 
the usual commutative spacetime,
\begin{equation}
f(x)=
\frac{1}{(2\pi)^{n/2}}\int d^4k~e^{ik_{\mu}x^{\mu}}\tilde{f}(k).
\end{equation}
This implements what is known as the Weyl-Moyal correspondence. This
relationship is preserved by the product of functions if we replace
the usual product of functions on commutative spacetime by the 
following Moyal-$\star$ product,
\begin{equation}
(f_1\star f_2)(x)=\left[\exp\left(\frac{i}{2}
\theta^{\mu\nu}\partial^x_{\mu}\partial^y_{\nu}\right)
f_1(x)f_2(y)\right]_{y=x}.
\end{equation}
Namely, using Eq. $(1)$ it is straightforward to show that 
$\hat{f}_1\hat{f}_2$ and $f_1\star f_2$ share the same Fourier 
tranform. In this sense we may study a field theory on NC spacetime 
by studying its counterpart on commutative spacetime with the usual 
product of functions replaced by the starred one. A different 
formalism based on the Seiberg-Witten map is developped in Refs. 
$\cite{wess}$. While there are problems such as the UV-IR mixing 
in the former approach, it is also not clear how to handle with the 
expansion of $\theta$ at quantum level in the latter $\cite{grosse}$. 
In this paper we will work in the former naive formalism. 

For convenience we list below some useful properties of the star 
product of functions which will be freely used in deriving Feynman 
rules.
\begin{equation}
\begin{array}{rcl}
(f_1\star f_2)\star f_3&=&f_1\star (f_2\star f_3),\\
(f_1\star f_2)^{\dagger}&=&f_2^{\dagger}\star f_1^{\dagger},\\
f_1\star f_2&=&f_2\star f_1|_{\theta\rightarrow -\theta},\\
\displaystyle\int d^nx~f_1\star f_2&=&
\displaystyle\int d^nx~f_2\star f_1,
\end{array}
\end{equation}
where the last one holds for functions which vanish fast enough at 
infinite spacetime.

{\it 2.2 Classical Lagrangian}

Instead of considering the general $U(N)$ gauge group, we restrict 
our explicit analysis to the $U(2)$ case. The reason is twofold. 
First, the $U(2)$ group is close to the standard model and thus 
physically well motivated. Second, it is algebraically easier to 
handle while it has a rich enough structure so that we expect the 
same conclusion should also be appropriate to the general $U(N)$ 
case.

We begin with the scalar potential which triggers the spontaneous
symmetry breakdown of $U(2)$ to $U(1)$,
\begin{equation}
V=-\mu^2\Phi^{\dagger}\star\Phi+
\lambda\Phi^{\dagger}\star\Phi\star\Phi^{\dagger}\star\Phi,
~~\mu^2>0,~~\lambda>0,
\end{equation}
where $\Phi$ is in the fundamental representation of $U(2)$. We 
assume its vacuum expection value is independent of $x$; without
loss of generality we take
\begin{equation}
\Phi=\phi+\phi_0,~~
\phi=\left(\begin{array}{c}\pi_+\\(\sigma+i\pi_0)/\sqrt{2}
           \end{array}
     \right),
\phi_0=\frac{v}{\sqrt{2}}\left(\begin{array}{c}0\\1
                               \end{array}
                         \right),
\end{equation}
with $v=\sqrt{\mu^2/\lambda}$. The negative potential in terms of 
shifted fields is
\begin{equation}
\begin{array}{rcl}
{\cal L}_{-V}&=&\displaystyle 
-\frac{1}{2}m_{\sigma}^2\sigma^2-\lambda\left\{
v\sigma(\sigma^2+\pi_0^2+2\pi_-\pi_+) +\pi_-\pi_+\pi_-\pi_+ 
+\frac{1}{4}(\sigma^4+\pi_0^4)\right.\\
&&\displaystyle\left.
+\pi_-\pi_+(\sigma^2+\pi_0^2) +\sigma^2\pi_0^2 
-\frac{1}{2}\sigma\pi_0\sigma\pi_0 +\pi_-\pi_+i[\sigma,\pi_0]
\right\},
\end{array}
\end{equation}
where $m_{\sigma}=v\sqrt{2\lambda}$ is the mass of the physical
Higgs boson. We have freely ignored terms which vanish upon spacetime 
integration using properties of the star product. We also suppress
the explicit $\star$ notation from now on.

It is convenient to formulate the gauge part by matrix. Denoting
\begin{equation}
t^A=\left\{
\begin{array}{ll}\displaystyle
                 \frac{1}{2}\sigma^A&~~{\rm for}~A=1,2,3\\
                 \displaystyle\frac{1}{2}1_2&~~{\rm for}~A=0
           \end{array}
    \right.,
{\rm ~with~~tr}(t^At^B)=\frac{1}{2}\delta^{AB},
\end{equation}
the gauge field is
\begin{equation}
G_{\mu}=G_{\mu}^At^A=\frac{1}{\sqrt{2}}\left(
\begin{array}{cc}A_{\mu}&W_{\mu}^+\\
                 W_{\mu}^-&Z_{\mu}
\end{array}
                                       \right).
\end{equation}
The Yang-Mills Lagrangian is
\begin{equation}
{\cal L}_G=-\frac{1}{2}{\rm Tr}~G_{\mu\nu}G^{\mu\nu},~~
G_{\mu\nu}=\partial_{\mu}G_{\nu}-\partial_{\nu}G_{\mu}
-ig[G_{\mu},G_{\nu}],
\end{equation}
where $g$ is the coupling. In terms of physical fields $A$, $Z$ and
$W^{\pm}$ as to be clear later on, we have
\begin{equation}
{\cal L}_G={\cal L}_{2G}+{\cal L}_{3G}+{\cal L}_{4G},
\end{equation}
where
\begin{equation}
\begin{array}{rcl}
{\cal L}_{2G}&=&\displaystyle 
-\frac{1}{4}(\partial_{\mu}A_{\nu}-\partial_{\nu}A_{\mu})^2
-\frac{1}{4}(\partial_{\mu}Z_{\nu}-\partial_{\nu}Z_{\mu})^2 \\
&&\displaystyle
-\frac{1}{2}(\partial_{\mu}W_{\nu}^+-\partial_{\nu}W_{\mu}^+)
(\partial^{\mu}W^{-\nu}-\partial^{\nu}W^{-\mu}),
\end{array}
\end{equation}
\begin{equation}
\begin{array}{rcl}
{\cal L}_{3G}&=&\displaystyle
+\frac{ig}{\sqrt{2}}\left(
\partial_{\mu}A_{\nu}[A^{\mu},A^{\nu}]+
\partial_{\mu}Z_{\nu}[Z^{\mu},Z^{\nu}]\right)\\
&&\displaystyle
+\frac{ig}{\sqrt{2}}A^{\mu}\left(
 W_{\nu}^+\partial_{\mu}W^{-\nu}+\partial^{\nu}W_{\mu}^+W^-_{\nu}
-\partial_{\mu}W_{\nu}^+W^{-\nu}-W_{\nu}^+\partial^{\nu}W^-_{\mu}
\right) \\
&&\displaystyle
+\frac{ig}{\sqrt{2}}Z^{\mu}\left(
 W_{\nu}^-\partial_{\mu}W^{+\nu}+\partial^{\nu}W_{\mu}^-W^+_{\nu}
-\partial_{\mu}W_{\nu}^-W^{+\nu}-W_{\nu}^-\partial^{\nu}W^+_{\mu}
\right) \\
&&\displaystyle
+\frac{ig}{\sqrt{2}}\left(
\partial^{\mu}A^{\nu}(W^+_{\mu}W^-_{\nu}-W^+_{\nu}W^-_{\mu})+
\partial^{\mu}Z^{\nu}(W^-_{\mu}W^+_{\nu}-W^-_{\nu}W^+_{\mu})
\right),
\end{array}
\end{equation}
\begin{equation}
\begin{array}{rcl}
{\cal L}_{4G}&=&\displaystyle
+\frac{g^2}{8}\left([A_{\mu},A_{\nu}]^2+[Z_{\mu},Z_{\nu}]^2\right)\\
&&\displaystyle
+\frac{g^2}{4}\left(2W^+_{\mu}W^-_{\nu}W^{+\mu}W^{-\nu}-
W^-_{\nu}W^{+\nu}W^-_{\mu}W^{+\mu}-W^+_{\nu}W^{-\nu}W^+_{\mu}W^{-\mu}
\right)\\
&&\displaystyle
+\frac{g^2}{2}\left([A^{\mu},A^{\nu}]W^+_{\mu}W^-_{\nu}
+A^{\mu}A^{\nu}W^+_{\mu}W^-_{\nu}-A^{\mu}A_{\mu}W^{+\nu}W^-_{\nu}
\right)\\
&&\displaystyle
+\frac{g^2}{2}\left([Z^{\mu},Z^{\nu}]W^-_{\mu}W^+_{\nu}
+Z^{\mu}Z^{\nu}W^-_{\mu}W^+_{\nu}-Z^{\mu}Z_{\mu}W^{-\nu}W^+_{\nu}
\right)\\
&&\displaystyle
+\frac{g^2}{2}A^{\mu}\left(2W^+_{\nu}Z_{\mu}W^{-\nu}
-W^+_{\nu}Z^{\nu}W^-_{\mu}-W^+_{\mu}Z^{\nu}W^-_{\nu}\right).
\end{array}
\end{equation}

The covariant kinetic Lagrangian for the scalar is
\begin{equation}
{\cal L}_{\Phi}=(D_{\mu}\Phi)^{\dagger}D^{\mu}\Phi,~~
D_{\mu}\Phi=\partial_{\mu}\Phi-igG_{\mu}\Phi.
\end{equation}
In terms of physical fields it can be cast in the form,
\begin{equation}
{\cal L}_{\Phi}={\cal L}_{2\phi}+{\cal L}_{G{\rm ~mass}}
+{\cal L}_{\phi G}+{\cal L}_{G\phi\phi}+{\cal L}_{GG\phi}
+{\cal L}_{GG\phi\phi},
\end{equation}
where
\begin{equation}
{\cal L}_{2\phi}=\frac{1}{2}(\partial_{\mu}\sigma)^2+
\frac{1}{2}(\partial_{\mu}\pi_0)^2
+\partial_{\mu}\pi_+\partial^{\mu}\pi_-,
\end{equation}
\begin{equation}
{\cal L}_{\rm G~mass}=\frac{1}{2}m_Z^2Z_{\mu}Z^{\mu}
+m_W^2W^+_{\mu}W^{-\mu},
\end{equation}
\begin{equation}
{\cal L}_{\phi G}=-m_ZZ^{\mu}\partial_{\mu}\pi_0
+im_W(W^-_{\mu}\partial^{\mu}\pi_+-W^+_{\mu}\partial^{\mu}\pi_-),
\end{equation}
\begin{equation}
\begin{array}{rcl}
{\cal L}_{G\phi\phi}&=&\displaystyle
+\frac{ig}{\sqrt{2}}A^{\mu}(\partial_{\mu}\pi_+\pi_-
-\pi_+\partial_{\mu}\pi_-)\\
&&\displaystyle
+\frac{ig}{2\sqrt{2}}Z^{\mu}\left(
 (\partial_{\mu}\sigma\sigma-\sigma\partial_{\mu}\sigma)
+(\partial_{\mu}\pi_0\pi_0-\pi_0\partial_{\mu}\pi_0)\right)\\
&&\displaystyle
+\frac{g}{2\sqrt{2}}Z^{\mu}\left(
 \pi_0\partial_{\mu}\sigma-\partial_{\mu}\pi_0\sigma
+\partial_{\mu}\sigma\pi_0-\sigma\partial_{\mu}\pi_0
\right)\\
&&\displaystyle
+\frac{ig}{2}W^+_{\mu}(\partial^{\mu}\sigma\pi_-
-\sigma\partial^{\mu}\pi_-)
+\frac{ig}{2}W^-_{\mu}(\partial^{\mu}\pi_+\sigma
-\pi_+\partial^{\mu}\sigma)\\
&&\displaystyle
+\frac{g}{2}W^+_{\mu}(\pi_0\partial^{\mu}\pi_-
-\partial^{\mu}\pi_0\pi_-)
+\frac{g}{2}W^-_{\mu}(\partial^{\mu}\pi_+\pi_0
-\pi_+\partial^{\mu}\pi_0),
\end{array}
\end{equation}
\begin{equation}
\begin{array}{rcl}
{\cal L}_{GG\phi}&=&\displaystyle
+\frac{g^2v}{2\sqrt{2}}\left(
 A^{\mu}(\pi_+W^-_{\mu} +W^+_{\mu}\pi_-)
+Z^{\mu}(W^-_{\mu}\pi_+ +\pi_-W^+_{\mu})\right)\\
&&\displaystyle
+\frac{g^2v}{2}(W^-_{\mu}W^{+\mu}+Z_{\mu}Z^{\mu})\sigma,
\end{array}
\end{equation}
\begin{equation}
\begin{array}{rcl}
{\cal L}_{GG\phi\phi}&=&
\displaystyle
+\frac{g^2}{2}\pi_+\pi_-\left(A_{\mu}A^{\mu}+W^+_{\mu}W^{-\mu}\right)
\\
&&\displaystyle
+\frac{g^2}{4}\left(\sigma^2+\pi_0^2+i[\pi_0,\sigma]\right)
\left(Z_{\mu}Z^{\mu}+W^-_{\mu}W^{+\mu}\right)
\\
&&\displaystyle
+\frac{g^2}{2\sqrt{2}}\sigma\left(
 (\pi_-A^{\mu}W^+_{\mu} +W^-_{\mu}A^{\mu}\pi_+)
+(\pi_-W^+_{\mu}Z^{\mu} +Z^{\mu}W^-_{\mu}\pi_+)\right)
\\
&&\displaystyle
+\frac{ig^2}{2\sqrt{2}}\pi_0\left(
 (\pi_-A^{\mu}W^+_{\mu}-W^-_{\mu}A^{\mu}\pi_+)
+(\pi_-W^+_{\mu}Z^{\mu}-Z^{\mu}W^-_{\mu}\pi_+)\right),
\end{array}
\end{equation}
and $m_W=gv/2$ and $m_Z=gv/\sqrt{2}$ are $W^{\pm}$ and $Z$
masses respectively.

The action defined by the classical Lagrangian
\begin{equation}
{\cal L}_{\rm class}=\displaystyle
{\cal L}_G+{\cal L}_{\Phi}+{\cal L}_{-V}
\end{equation}
is invariant under the generalized, starred $U(2)$ transformation,
\begin{equation}
\begin{array}{rcl}
G_{\mu}&\to & G^{\prime}_{\mu}=U\star G_{\mu}\star U^{-1}
+ig^{-1}U\star\partial_{\mu}U^{-1},\\
\Phi &\to & \Phi^{\prime}=U\star\Phi,
\end{array}
\end{equation}
where $U=\exp(ig\eta(x))_{\star}$ and we have restored the 
explicit star notation for clearness.

{\it 2.3 Gauge fixing and ghost terms}

To make the theory well-defined and to quantize it, we should do
gauge fixing and include its corresponding ghost terms. Since the
quadratic terms in the action remain the same on NC spacetime, it
is easy to expect how to generalize the gauge fixing procedure;
namely we replace the usual product by the starred one (again
suppressing the notation from now on),
\begin{equation}
\begin{array}{rcl}
{\cal L}_{\rm g.f.}&=&\displaystyle
-\frac{1}{\xi}{\rm Tr}(ff),\\
f&=&\partial^{\mu}G_{\mu}+ig\xi(\phi^{\dagger}t^A\phi_0
-\phi_0^{\dagger}t^A\phi)t^A.
\end{array}
\end{equation}
To construct the ghost terms we first generalize the BRS
transformation to NC spacetime,
\begin{equation}
\begin{array}{rcl}
\delta G_{\mu}&=&\epsilon(\partial_{\mu}c+ig[c,G_{\mu}]),\\
\delta\phi&=&\epsilon igc\Phi,~~
\delta\phi^{\dagger}=-\epsilon ig\Phi^{\dagger}c,\\
\delta c&=&\epsilon igcc,
\end{array}
\end{equation}
where $\epsilon$ is an infinitesimal Grassmann constant and $c$ is
the ghost field. We have
\begin{equation}
{\cal L}_{\rm ghost}=-2{\rm Tr}(\bar{c}{\sf s}f),
\end{equation}
where $\epsilon {\sf s}f$ is the BRS transformation of the gauge
fixing function $f$ and thus,
\begin{equation}
{\sf s}f=\partial^{\mu}(\partial_{\mu}c+ig[c,G_{\mu}])
+g^2\xi(\Phi^{\dagger}ct^A\phi_0+\phi_0^{\dagger}t^Ac\Phi)t^A.
\end{equation}
Noting that ${\sf s}^2f=0$, the BRS invariance of the sum 
${\cal L}_{\rm g.f.}+{\cal L}_{\rm ghost}$ is then guaranteed by
requiring
\begin{equation}
\delta\bar{c}=-\frac{1}{\xi}\epsilon f.
\end{equation}
We parametrize the ghost fields as follows,
\begin{equation}
\begin{array}{l}
\displaystyle
c=\frac{1}{\sqrt{2}}\left(
\begin{array}{cc}c_A&c_+\\c_-&c_Z
\end{array}
                    \right),~~
\bar{c}=\frac{1}{\sqrt{2}}\left(
\begin{array}{cc}\bar{c}_A&\bar{c}_+\\\bar{c}_-&\bar{c}_Z
\end{array}
                          \right).
\end{array}
\end{equation}
Then, the explicit forms of 
${\cal L}_{\rm g.f.}$ and ${\cal L}_{\rm ghost}$
are,
\begin{equation}
\begin{array}{rcl}
{\cal L}_{\rm g.f.}&=&\displaystyle
-\frac{1}{2\xi}\left(
(\partial^{\mu}A_{\mu})^2+(\partial^{\mu}Z_{\mu})^2\right)
-\frac{1}{\xi}\partial^{\mu}W^+_{\mu}\partial^{\nu}W^-_{\nu}\\
&&\displaystyle -\frac{1}{2}\xi m^2_Z\pi^2_0-\xi m_W^2\pi_+\pi_-\\
&&\displaystyle -m_Z\pi_0\partial^{\mu}Z_{\mu}
-im_W(\pi_-\partial^{\mu}W^+_{\mu}-\pi_+\partial^{\mu}W^-_{\mu}),\\
{\cal L}_{\rm ghost}&=&
{\cal L}_{c\bar{c}}+{\cal L}_{\phi c\bar{c}}+{\cal L}_{G c\bar{c}},
\end{array}
\end{equation}
\begin{equation}
\begin{array}{rcl}
{\cal L}_{c\bar{c}}&=&\displaystyle
-\bar{c}_A\partial^2c_A-\bar{c}_Z(\partial^2+\xi m_Z^2)c_Z
-\bar{c}_-(\partial^2+\xi m_W^2)c_+
-\bar{c}_+(\partial^2+\xi m_W^2)c_-,\\
\end{array}
\end{equation}
\begin{equation}
\begin{array}{rcl}
{\cal L}_{\phi c\bar{c}}&=&\displaystyle
-\frac{1}{4}\xi g^2v\bar{c}_-\left(
\sqrt{2}c_A\pi_+ +c_+(\sigma+i\pi_0)\right)\\
&&\displaystyle
-\frac{1}{4}\xi g^2v\bar{c}_+\left(
\sqrt{2}\pi_-c_A +(\sigma-i\pi_0)c_-\right)
\\
&&\displaystyle
-\frac{1}{4}\xi g^2v\bar{c}_Z\left(
\sqrt{2}(\pi_-c_++c_-\pi_+)+\{\sigma,c_Z\}+i[c_Z,\pi_0]\right),
\end{array}
\end{equation}
\begin{equation}
\begin{array}{rcl}
{\cal L}_{G c\bar{c}}&=&\displaystyle
-\frac{ig}{\sqrt{2}}\bar{c}_A\partial^{\mu}\left(
[c_A,A_{\mu}]+(c_+W^-_{\mu}-W^+_{\mu}c_-)\right)\\
&&\displaystyle
-\frac{ig}{\sqrt{2}}\bar{c}_Z\partial^{\mu}\left(
[c_Z,Z_{\mu}]+(c_-W^+_{\mu}-W^-_{\mu}c_+)\right)\\
&&\displaystyle
-\frac{ig}{\sqrt{2}}\bar{c}_-\partial^{\mu}\left(
c_A W^+_{\mu}-A_{\mu}c_+ +c_+Z_{\mu}-W^+_{\mu}c_Z\right)\\
&&\displaystyle
-\frac{ig}{\sqrt{2}}\bar{c}_+\partial^{\mu}\left(
c_-A_{\mu} -W^-_{\mu}c_A +c_ZW^-_{\mu} -Z_{\mu}c_-\right).
\end{array}
\end{equation}
Note that the $G\phi$ mixing terms in ${\cal L}_{\rm g.f.}$ are
cancelled by ${\cal L}_{\phi G}$. The complete Feynman rules are
collected in Appendix A.

{\it 2.4 Renormalization constants and counterterms}

To go beyond the lowest order we introduce the following 
renormalization constants for bare quantities,
\begin{equation}
\begin{array}{l}
(G_{\mu})_{\rm B}=Z^{1/2}_GG_{\mu},~~
(\phi)_{\rm B}=Z^{1/2}_{\phi}\phi,\\
\displaystyle (g)_{\rm B}=Z^{-1/2}_GZ_gg,~~
(\lambda)_{\rm B}=Z^{-2}_{\phi}Z_{\lambda}\lambda,\\
\displaystyle (\mu^2)_{\rm B}=Z^{-1}_{\phi}\mu^2\left(
1+\frac{\delta\mu^2}{\mu^2}\right),~~
(v)_{\rm B}=Z^{1/2}_{\phi}v\left(1+\frac{\delta v}{v}\right).
\end{array}
\end{equation}
Note that the redundant renormalization constant $\delta v$ will
be determined by the additional requirement that the $\sigma$ 
tadpole be cancelled at one loop. We have chosen the same 
renormalization constant for all members of a multiplet. As to be
shown below this will be sufficent for removing divergences.

The counterterms introduced by the above substitutions are standard
for the linear and quadratic terms since no difference arises as
compared to the commutative case. These are listed explicitly in 
Appendeix A together with the rules to obtain counterterms for 
vertices in ${\cal L}_{\rm class}$. We focus below on deriving 
counterterms in the gauge fixing sector. 

Since the gauge fixing function can be chosen at will, we 
choose it to be given in terms of renormalized quantities,
\begin{equation}
(\xi)_{\rm B}=\xi,~~(f)_{\rm B}=f=
\partial^{\mu}G_{\mu}+\frac{igv\xi}{\sqrt{2}}\left(\phi^{\dagger}
t^A\hat{\phi}_0-\hat{\phi}_0^{\dagger}t^A\phi\right)t^A,
\end{equation}
where $\hat{\phi}_0^{\dagger}=(0~~1)$.
We require that 
$(D_{\mu}\Phi)_{\rm B}=Z^{1/2}_{\phi}(\partial_{\mu}-igZ_gG_{\mu})
\Phi$ be covariant under the infinitesimal gauge transformation for
renormalized fields,
\begin{equation}
\begin{array}{rcl}
\delta G_{\mu}&=&y\partial_{\mu}\eta+zig[\eta,G_{\mu}],\\
\delta\Phi&=&xig\eta\Phi,~~
\delta\Phi^{\dagger}=-xig\Phi^{\dagger}\eta.
\end{array}
\end{equation}
This determines the constants $x=z=Z_gy$. Now consider the BRS
transformation for renormalized fields,
\begin{equation}
\begin{array}{rcl}
\delta G_{\mu}&=&\epsilon(y\partial_{\mu}c+zig[c,G_{\mu}]),\\
\delta\phi&=&\epsilon xigc\Phi,~~
\delta\phi^{\dagger}=-\epsilon xig\Phi^{\dagger}c,\\
\delta c&=&\epsilon uigcc.
\end{array}
\end{equation}
We find,
\begin{equation}
{\sf s}f=\partial^{\mu}(y\partial_{\mu}c+zig[c,G_{\mu}])
+xg^2v\xi/\sqrt{2}(\Phi^{\dagger}ct^A\hat{\phi}_0
+\hat{\phi}_0^{\dagger}t^Ac\Phi)t^A.
\end{equation}
Note that $\Phi$ now contains $v$ in the form of $v+\delta v$.
We hope to keep Eq. $(30)$ intact so that 
$(\bar{c})_{\rm B}=\bar{c}$. The invariance of the sum 
${\cal L}_{\rm g.f.}+{\cal L}_{\rm ghost}$ under BRS 
transformations for renormalized fields is again guaranteed by
the nilpotency ${\sf s}^2f=0$. This then implies $x=z=u$.
Now we introduce the field renormalization constant for ghosts,
\begin{equation}
(c)_{\rm B}=Z_cc.
\end{equation}
Since the first term in ${\sf s}f$ gives the ghost kinetic terms,
it is natural to identify $y=Z_c$. This fixes all constants in
${\sf s}f$,
\begin{equation}
\begin{array}{rcl}
{\sf s}f&=&Z_c\partial^{\mu}(\partial_{\mu}c+igZ_g[c,G_{\mu}])\\
&&+Z_cZ_g(1+\delta v/v)g^2v^2\xi/2
(\hat{\phi}_0^{\dagger}ct^A\hat{\phi}_0
+\hat{\phi}_0^{\dagger}t^Ac\hat{\phi}_0)t^A \\
&&+Z_cZ_gg^2v\xi/\sqrt{2}
(\phi^{\dagger}ct^A\hat{\phi}_0
+\hat{\phi}_0^{\dagger}t^Ac\phi)t^A.
\end{array}
\end{equation}
The counterterms for self-energies and the rules for vertices in
the gauge fixing sector are also included in Appendix A.

\begin{center}
{\bf 3. One loop divergences and renormalization constants}
\end{center}

In this section we present our results of one loop divergences in
1PI Green's functions. Although we have exhausted all possibilities
for each type of functions discussed below, it is not possible and 
also unnecessary to list all of them. Actually, a glance at the 
counterterms shows that all functions in the same type must have 
the same or similar divergent structure if divergences can be 
removed altogether with the renormalization constants introduced 
above. Instead, we demonstrate our results by typical examples. 
Since we are interested in the ultraviolet (UV) divergences, 
only diagrams which are apparently divergent by power counting are 
computed below. The complete Feynman diagrams are shown in 
Appendix B. We thus will not touch upon the UV-IR mixing problem for 
exceptional momenta such as $\theta_{\mu\nu}p^{\nu}=0$, etc. We 
work in the $\xi=1$ gauge throughout for simplicity. Then the 
would-be Goldstone bosons and ghosts have the same masses as their 
corresponding gauge bosons.

{\it 3.1 Tadpole}

The Feynman diagrams are shown in Fig. $1$. The result is
\begin{equation}
\begin{array}{rcl}
iT&=&\displaystyle 
+\lambda v\int\left[3D^{\sigma}_k+D^Z_k+2D^W_k\right]\\
&&\displaystyle
+\frac{n}{2}g^2v\int\left[D^W_k+D^Z_k\right]\\
&&\displaystyle
-\frac{1}{2}g^2v\int\left[D^W_k+D^Z_k\right],
\end{array}
\end{equation}
where
\begin{equation}
\int=\int\frac{d^nk}{(2\pi)^n},~~D^j_p=\frac{1}{p^2-m^2_j+i\epsilon}.
\end{equation}
We work in $n=4-2\epsilon$ dimensions to regularize the UV 
divergences. Note that there is no $\theta$ dependence in
$iT$ since we need at least two independent momenta for this.
We obtain the divergent part as follows,
\begin{equation}
iT=i\deltae\left[6\lambda^2+\lambda g^2+\frac{9}{8}g^4\right]v^3,
~~\deltae=\frac{1}{(4\pi)^2}\frac{1}{\epsilon}.
\end{equation}
\begin{center}
\begin{picture}(300,90)(0,0)
\DashCArc(30,50)(20,0,360){5}
\DashLine(50,50)(80,50){5}
\Text(30,75)[]{$\sigma,\pi_0,\pi_+$}
\Text(75,55)[]{$\sigma$}
\SetOffset(100,0)
\PhotonArc(30,50)(20,0,360){2}{12}
\DashLine(50,50)(80,50){5}
\Text(30,75)[]{\scriptsize$W^+,Z$}
\SetOffset(200,0)
\DashCArc(30,50)(20,0,360){1}
\DashLine(50,50)(80,50){5}
\Text(30,75)[]{$c_{\pm},c_Z$}
\SetOffset(0,0)
\Text(150,10)[]{Figure $1$. $\sigma$ tadpole}
\end{picture}\\
\end{center}

{\it 3.2 $\phi\phi$ self-energies and mixings}

We have divergent $\sigma\sigma$, $\pi_0\pi_0$, $\pi_+\pi_-$
self-energies and the finite $\sigma\pi_0$ mixing. We compute
the first and the last ones as examples. The Feynman diagrams
for $\sigma\sigma$ are shown in Fig. $2$. The result is
\begin{equation}
\begin{array}{rcl}
i\Sigma^{\sigma\sigma}(p)&=&\displaystyle
+\lambda^2 v^2\int\left[c^2_{k\wedge p}\left(
18D^{\sigma}_k D^{\sigma}_{k+p}+2D^Z_kD^Z_{k+p}
\right)+4D^W_k D^W_{k+p}\right]\\
&&\displaystyle 
-\frac{1}{2}g^2\int(k+2p)^2\left[
s^2_{k\wedge p}D^Z_kD^{\sigma}_{k+p}+
c^2_{k\wedge p}D^Z_kD^Z_{k+p}+
D^W_kD^W_{k+p}\right]
\\
&&\displaystyle 
+\frac{n}{4}g^4v^2\int\left[
2c^2_{k\wedge p}D^Z_kD^Z_{k+p}+D^W_kD^W_{k+p}\right]
\\
&&\displaystyle 
-\frac{1}{8}g^4v^2\int\left[
2c^2_{k\wedge p}D^Z_kD^Z_{k+p}+D^W_kD^W_{k+p}\right]
\\
&&\displaystyle 
+\lambda\int\left[(1+2c^2_{k\wedge p})D^{\sigma}_k+
(2-c_{k\wedge p})D^Z_{k}+2D^W_k\right]\\
&&\displaystyle
+\frac{n}{2}g^2\int\left[D^W_k+D^Z_k\right],
\end{array}
\end{equation}
where 
$s^m_{k\wedge p}=\sin^m(k\wedge p)$,
$c^m_{k\wedge p}=\cos^m(k\wedge p)$ and 
$k\wedge p=\theta_{\mu\nu}k^{\mu}p^{\nu}/2$. 
Using 
$s^2_{k\wedge p}=(1-c_{2k\wedge p})/2$ and
$c^2_{k\wedge p}=(1+c_{2k\wedge p})/2$ to separate the planar
from the nonplanar part, and noting that the latter is finite
due to the oscillating factor, we can isolate the UV divergences
and obtain,
\begin{equation}
i\Sigma^{\sigma\sigma}(p)=i\deltae\left[
-2g^2p^2+18\lambda^2 v^2+\lambda g^2v^2+\frac{21}{8}g^4v^2\right].
\end{equation}
A similar calculation gives,
\begin{equation}
i\Sigma^{\pi_0\pi_0}(p)=i\Sigma^{\pi_+\pi_-}(p)=
i\deltae\left[
-2g^2p^2+6\lambda^2 v^2+\lambda g^2v^2+\frac{9}{8}g^4v^2\right].
\end{equation}
\begin{center}
\begin{picture}(400,180)(0,0)
\SetOffset(0,90)
\DashCArc(50,50)(20,0,360){5}
\DashLine(10,50)(30,50){5}\DashLine(70,50)(90,50){5}
\Text(50,75)[]{$\sigma,\pi_0,\pi_+$}
\Text(50,25)[]{$\sigma,\pi_0,\pi_+$}
\Text(50,62)[]{$k$}
\Text(15,55)[]{$\sigma$}\Text(20,42)[]{$p$}
\Text(85,55)[]{$\sigma$}
\SetOffset(100,90)
\PhotonArc(50,50)(20,0,180){2}{6}\DashCArc(50,50)(20,180,360){5}
\DashLine(10,50)(30,50){5}\DashLine(70,50)(90,50){5}
\Text(50,75)[]{\scriptsize$Z;W^{\pm}$}
\Text(50,25)[]{$\sigma,\pi_0;\pi_{\pm}$}
\SetOffset(200,90)
\PhotonArc(50,50)(20,0,360){2}{12}
\DashLine(10,50)(30,50){5}\DashLine(70,50)(90,50){5}
\Text(50,75)[]{\scriptsize$Z,W^+$}
\Text(50,25)[]{\scriptsize$Z,W^+$}
\SetOffset(300,90)
\DashCArc(50,50)(20,0,360){1}
\DashLine(10,50)(30,50){5}\DashLine(70,50)(90,50){5}
\Text(50,75)[]{$c_Z,c_{\pm}$}
\Text(50,25)[]{$c_Z,c_{\pm}$}
\SetOffset(0,0)
\DashCArc(50,60)(20,0,360){5}
\DashLine(10,40)(90,40){5}
\Text(50,85)[]{$\sigma,\pi_0,\pi_+$}
\SetOffset(100,0)
\PhotonArc(50,62)(20,-90,270){2}{12}
\DashLine(10,40)(90,40){5}
\Text(50,87)[]{\scriptsize$Z,W^+$}
\SetOffset(0,0)
\Text(200,10)[]{Figure $2$. $\sigma$ self-energy}
\end{picture}\\
\end{center}

Now we come to the $\sigma\pi_0$ mixing shown in Fig. $3$. All of
the neutral particle loops are found to be proportional to 
$s_{2k\wedge p}$ and thus finite. The $W^+\pi_+$ loop cancels
exactly the $W^-\pi_-$ loop due to the sign flip in the $\sigma$
couplings although they are separately divergent after 
cancellation of oscillating factors from the two vertices.
The same happens for the $c_{\pm}$ loops due to the sign flip in 
the $\pi_0$ vertices. Thus we finally have a finite 
$\sigma\pi_0$ mixing.
\begin{center}
\begin{picture}(300,90)(0,0)
\SetOffset(0,10)
\PhotonArc(50,50)(20,0,180){2}{6}\DashCArc(50,50)(20,180,360){5}
\DashLine(10,50)(30,50){5}\DashLine(70,50)(90,50){5}
\Text(50,75)[]{\scriptsize$Z;W^{\pm}$}
\Text(50,25)[]{$\sigma,\pi_0;\pi_{\pm}$}
\Text(50,62)[]{$k$}
\Text(15,55)[]{$\sigma$}\Text(20,42)[]{$p$}
\Text(85,55)[]{$\pi_0$}
\SetOffset(150,10)
\DashCArc(50,50)(20,0,360){1}
\DashLine(10,50)(30,50){5}\DashLine(70,50)(90,50){5}
\Text(50,75)[]{$c_Z,c_{\pm}$}
\Text(50,25)[]{$c_Z,c_{\pm}$}
\SetOffset(0,0)
\Text(150,10)[]{Figure $3$. $\sigma\pi_0$ mixing}
\end{picture}\\
\end{center}

{\it 3.3 $GG$ self-energies and mixings}

We have divergent $AA$, $ZZ$ and $W^+W^-$ self-energies and a 
finite $AZ$ mixing. For the divergent one we take as an example 
the $A$ self-energy whose diagrams are shown in Fig. $4$. 
We find,
\begin{equation}
\begin{array}{rcl}
i\Sigma^{AA}_{\mu\nu}(p)&=&\displaystyle
+\frac{1}{2}g^2\int(2k+p)_{\mu}(2k+p)_{\nu}D^W_kD^W_{k+p}
-\frac{1}{4}g^4v^2\gmunu\int D^W_kD^W_{k+p}
\\
&&\displaystyle
+\frac{1}{2}g^2\int P^2_{\mu\nu}
\left[2s^2_{k\wedge p}D^A_kD^A_{k+p}+D^W_kD^W_{k+p}\right]
\\
&&\displaystyle
-g^2\int(k+p)_{\mu}k_{\nu}\left[2s^2_{k\wedge p}D^A_kD^A_{k+p}
+D^W_kD^W_{k+p}\right]
\\
&&\displaystyle
-g^2\gmunu\int D^W_k
+g^2(1-n)\gmunu\int\left[2s^2_{k\wedge p}D^A_k+D^W_k\right]\\
&=&\displaystyle
i\deltae\frac{19}{6}g^2(p^2\gmunu-p_{\mu}p_{\nu}),
\end{array}
\end{equation}
where $P^2_{\mu\nu}=P_{\alpha\beta\mu}(k,-k-p,p)
P^{\alpha\beta}_{~~~\nu}(-k,k+p,-p)$ with
$P_{\alpha\beta\gamma}(k_1,k_2,k_3)$ given in Appendix A.
Identical results are obtained for the other two self-energies,
especially there are no divergences proportional to $m^2_{W,Z}$.
\begin{center}
\begin{picture}(400,180)(0,0)
\SetOffset(0,90)
\DashCArc(50,50)(20,0,360){5}
\Photon(10,50)(30,50){3}{3}\Photon(70,50)(90,50){3}{3}
\Text(50,75)[]{$\pi_+$}
\Text(50,25)[]{$\pi_+$}
\Text(50,62)[]{$k$}
\Text(15,58)[]{\scriptsize$A_{\mu}$}\Text(20,42)[]{$p$}
\Text(85,58)[]{\scriptsize$A_{\nu}$}
\SetOffset(100,90)
\PhotonArc(50,50)(20,0,180){2}{6}\DashCArc(50,50)(20,180,360){5}
\Photon(10,50)(30,50){3}{3}\Photon(70,50)(90,50){3}{3}
\Text(50,75)[]{\scriptsize$W^{\pm}$}
\Text(50,25)[]{$\pi_{\pm}$}
\Text(50,75)[]{\scriptsize$W^{\pm}$}
\Text(50,25)[]{$\pi_{\pm}$}
\SetOffset(200,90)
\PhotonArc(50,50)(20,0,360){2}{12}
\Photon(10,50)(30,50){3}{3}\Photon(70,50)(90,50){3}{3}
\Text(50,75)[]{\scriptsize$A,W^+$}
\Text(50,25)[]{\scriptsize$A,W^+$}
\SetOffset(300,90)
\DashCArc(50,50)(20,0,360){1}
\Photon(10,50)(30,50){3}{3}\Photon(70,50)(90,50){3}{3}
\Text(50,75)[]{$c_A,c_{\pm}$}
\Text(50,25)[]{$c_A,c_{\pm}$}
\SetOffset(0,0)
\DashCArc(50,62)(20,0,360){5}
\Photon(10,40)(90,40){3}{10}
\Text(50,85)[]{$\pi_+$}
\SetOffset(100,0)
\PhotonArc(50,64)(20,-90,270){2}{12}
\Photon(10,40)(90,40){3}{10}
\Text(50,90)[]{\scriptsize$A,W^+$}
\SetOffset(0,0)
\Text(200,10)[]{Figure $4$. $A$ self-energy}
\end{picture}\\
\end{center}

The Feynman diagrams for the $AZ$ mixing are shown in Fig. $5$.
Each diagram is finite due to an oscillating factor 
$\exp(i2k\wedge p)$. This occurs because the $A$ and $Z$
couplings to the same charged particles have opposite phases
which become the same when the charges are reversed in one of
the couplings.
\begin{center}
\begin{picture}(400,90)(0,0)
\SetOffset(0,10)
\PhotonArc(50,50)(20,0,180){2}{6}\DashCArc(50,50)(20,180,360){5}
\Photon(10,50)(30,50){3}{3}\Photon(70,50)(90,50){3}{3}
\Text(50,75)[]{\scriptsize$W^{\pm}$}
\Text(50,25)[]{$\pi_{\pm}$}
\Text(50,62)[]{$k$}
\Text(15,58)[]{\scriptsize$A_{\mu}$}\Text(20,42)[]{$p$}
\Text(85,58)[]{\scriptsize$Z_{\nu}$}
\SetOffset(100,10)
\PhotonArc(50,50)(20,0,360){2}{12}
\Photon(10,50)(30,50){3}{3}\Photon(70,50)(90,50){3}{3}
\Text(50,75)[]{\scriptsize$W^+$}
\Text(50,25)[]{\scriptsize$W^+$}
\SetOffset(200,10)
\DashCArc(50,50)(20,0,360){1}
\Photon(10,50)(30,50){3}{3}\Photon(70,50)(90,50){3}{3}
\Text(50,75)[]{$c_{\pm}$}
\Text(50,25)[]{$c_{\pm}$}
\SetOffset(300,0)
\PhotonArc(50,64)(20,-90,270){2}{12}
\Photon(10,40)(90,40){3}{10}
\Text(50,90)[]{\scriptsize$W^+$}
\SetOffset(0,0)
\Text(200,10)[]{Figure $5$. $AZ$ mixing}
\end{picture}\\
\end{center}

{\it 3.4 $G\phi$ mixings}

We have divergent $Z\pi_0$ and $W^{\pm}\pi_{\mp}$ mixings while
$A\sigma$, $A\pi_0$ and $Z\sigma$ must be finite. The diagrams
for $W^+\pi_-$ are shown in Fig. $6$ where $p$ denotes the incoming
momentum of $W^+$. We obtain,
\begin{equation}
\begin{array}{rcl}
i\Sigma^{W^+\pi_-}_{\mu}(p)&=&\displaystyle
-gv\lambda\int(2k+p)_{\mu}D^{\sigma}_kD^W_{k+p}
\\
&&\displaystyle
-\frac{1}{4}g^3v\int(k+2p)_{\mu}\left[
D^A_kD^W_{k+p}+D^W_kD^{\sigma}_{k+p}\right]
\\
&&\displaystyle
+\frac{1}{4}g^3v(1-n)\int(2k+p)_{\mu}\left[
D^W_kD^A_{k+p}-D^W_kD^Z_{k+p}\right]
\\
&&\displaystyle
+\frac{1}{4}g^3v\int(k+p)_{\mu}\left[
D^W_kD^A_{k+p}+D^Z_kD^W_{k+p}\right]
\\
&=&\displaystyle
i\deltae (-g^2m_W)p_{\mu}.
\end{array}
\end{equation}
\begin{center}
\begin{picture}(400,90)(0,0)
\SetOffset(0,10)
\DashCArc(50,50)(20,0,360){5}
\Photon(10,50)(30,50){3}{3}\DashLine(70,50)(90,50){5}
\Text(50,75)[]{$\sigma$}
\Text(50,25)[]{$\pi_+$}
\Text(50,62)[]{$k$}
\Text(15,58)[]{\scriptsize$W^+_{\mu}$}\Text(20,42)[]{$p$}
\Text(85,58)[]{$\pi_-$}
\SetOffset(100,10)
\PhotonArc(50,50)(20,0,180){2}{6}\DashCArc(50,50)(20,180,360){5}
\Photon(10,50)(30,50){3}{3}\DashLine(70,50)(90,50){5}
\Text(50,75)[]{\scriptsize$A,W^-$}
\Text(50,25)[]{$\pi_+,\sigma$}
\SetOffset(200,10)
\PhotonArc(50,50)(20,0,360){2}{12}
\Photon(10,50)(30,50){3}{3}\DashLine(70,50)(90,50){5}
\Text(50,75)[]{\scriptsize$W^-$}
\Text(50,25)[]{\scriptsize$A,Z$}
\SetOffset(300,10)
\DashCArc(50,50)(20,0,360){1}
\Photon(10,50)(30,50){3}{3}\DashLine(70,50)(90,50){5}
\Text(50,75)[]{$c_-,c_Z$}
\Text(50,25)[]{$c_A,c_+$}
\SetOffset(0,0)
\Text(200,10)[]{Figure $6$. $W^+\pi_-$ mixing}
\end{picture}\\
\end{center}

It is interesting to see how the $Z\sigma$ mixing shown in 
Fig. $7$ becomes finite. The situation is similar to the case of
the $\sigma\pi_0$ mixing. All neutral particle loops are
proportional to $s_{2k\wedge p}$ and thus finite. The 
$W^{\pm}\pi_{\pm}$ loops exactly cancel themselves because of
a sign flip in their couplings to $\sigma$. The same is true
for the $c_{\pm}$ loops but this time the flip
occurs in the $Z$ couplings. Finally, the $W^{\pm}$ loop is 
finite since it is simply proportional to $(2k+p)_{\mu}$.
\begin{center}
\begin{picture}(400,90)(0,0)
\SetOffset(0,10)
\DashCArc(50,50)(20,0,360){5}
\Photon(10,50)(30,50){3}{3}\DashLine(70,50)(90,50){5}
\Text(50,75)[]{$\sigma,\pi_0$}
\Text(50,25)[]{$\sigma,\pi_0$}
\Text(50,62)[]{$k$}
\Text(15,58)[]{\scriptsize$Z_{\mu}$}\Text(20,42)[]{$p$}
\Text(85,58)[]{$\sigma$}
\SetOffset(100,10)
\PhotonArc(50,50)(20,0,180){2}{6}\DashCArc(50,50)(20,180,360){5}
\Photon(10,50)(30,50){3}{3}\DashLine(70,50)(90,50){5}
\Text(50,75)[]{\scriptsize$Z,W^{\pm}$}
\Text(50,25)[]{$\sigma,\pi_{\pm}$}
\SetOffset(200,10)
\PhotonArc(50,50)(20,0,360){2}{12}
\Photon(10,50)(30,50){3}{3}\DashLine(70,50)(90,50){5}
\Text(50,75)[]{\scriptsize$Z,W^+$}
\Text(50,25)[]{\scriptsize$Z,W^+$}
\SetOffset(300,10)
\DashCArc(50,50)(20,0,360){1}
\Photon(10,50)(30,50){3}{3}\DashLine(70,50)(90,50){5}
\Text(50,75)[]{$c_Z,c_{\pm}$}
\Text(50,25)[]{$c_Z,c_{\pm}$}
\SetOffset(0,0)
\Text(200,10)[]{Figure $7$. $Z\sigma$ mixing}
\end{picture}\\
\end{center}

{\it 3.5 $c\bar{c}$ self-energies and mixings}

The $c\bar{c}$ self-energies are the easiest to compute. We
have divergent combinations
$c_A\bar{c}_A$, $c_Z\bar{c}_Z$  and $c_{\pm}\bar{c}_{\mp}$, 
and the finite ones,
$c_A\bar{c}_Z$ and $c_Z\bar{c}_A$. The result for the 
$c_+\bar{c}_-$ self-energy shown in Fig. $8$ is,
\begin{equation}
\begin{array}{rcl}
i\Sigma^{c_+\bar{c}_-}(p)&=&\displaystyle
+\frac{1}{16}g^4v^2\int D^W_k\left[D^{\sigma}_{k+p}-D^Z_{k+p}
\right]
\\
&&\displaystyle
+\frac{1}{2}g^2\int k\cdot p \left[D^W_k\left(D^A_{k+p}+D^Z_{k+p}
\right)+\left(D^A_k+D^Z_k\right)D^W_{k+p}\right]
\\
&=&\displaystyle i\deltae(-g^2p^2).
\end{array}
\end{equation}
Note that there is no divergence corresponding to the mass term. 
The $c_A\bar{c}_Z$ mixing shown in Fig. $9$ is finite since each
diagram contains an oscillating factor $\exp(\pm i2k\wedge p)$.
\begin{center}
\begin{picture}(300,80)(0,0)
\DashCArc(50,40)(20,0,180){5}
\DashLine(10,40)(30,40){1}
\DashArrowLine(30,40)(70,40){1}
\DashLine(90,40)(70,40){1}
\Text(50,65)[]{$\sigma,\pi_0$}
\Text(50,32)[]{$c_+$}
\Text(50,45)[]{$-k$}
\Text(15,32)[]{$c_+$}\Text(20,45)[]{$p$}
\Text(85,32)[]{$\bar{c}_-$}
\SetOffset(150,0)
\PhotonArc(50,40)(20,0,180){2}{6}
\DashLine(10,40)(30,40){1}
\DashArrowLine(30,40)(70,40){1}
\DashLine(90,40)(70,40){1}
\Text(50,65)[]{\scriptsize$A,Z;W^-$}
\Text(50,32)[]{$c_+;c_A,c_Z$}
\SetOffset(0,0)
\Text(150,10)[]{Figure $8$. $c_+\bar{c}_-$ self-energy}
\end{picture}\\
\end{center}
\begin{center}
\begin{picture}(300,80)(0,0)
\DashCArc(50,40)(20,0,180){5}
\DashLine(10,40)(30,40){1}
\DashArrowLine(30,40)(70,40){1}
\DashLine(90,40)(70,40){1}
\Text(50,65)[]{$\pi_{\pm}$}
\Text(50,32)[]{$c_{\pm}$}
\Text(50,45)[]{$-k$}
\Text(15,32)[]{$c_A$}\Text(20,45)[]{$p$}
\Text(85,32)[]{$\bar{c}_Z$}
\SetOffset(150,0)
\PhotonArc(50,40)(20,0,180){2}{6}
\DashLine(10,40)(30,40){1}
\DashArrowLine(30,40)(70,40){1}
\DashLine(90,40)(70,40){1}
\Text(50,65)[]{\scriptsize$W^{\pm}$}
\Text(50,32)[]{$c_{\pm}$}
\SetOffset(0,0)
\Text(150,10)[]{Figure $9$. $c_A\bar{c}_Z$ mixing}
\end{picture}\\
\end{center}

{\it 3.6 $\phi\phi\phi$ vertices}

Now we start our computation of vertices with the trilinear scalar
couplings. We have $\sigma\sigma\sigma$, $\sigma\pi_0\pi_0$ and 
$\sigma\pi_+\pi_-$ vertices which may be divergent, and
$\sigma\sigma\pi_0$, $\pi_0\pi_0\pi_0$ and $\pi_0\pi_+\pi_-$ 
vertices which must be finite. We illustrate the computation with
the last one of each type whose diagrams are shown in Figs. 10 and 11
respectively. The arrow inside the loop indicates the charge flow. 
$p$ and $p_{\pm}$ are the incoming momenta of $\sigma$ or $\pi_0$, 
and $\pi_{\pm}$. 

Let us first consider the $\sigma\pi_+\pi_-$ vertex. Fig. $(a)$ 
contains the phase factor of 
$\exp(\pm i2p\wedge k)$
and is thus finite. The same is true for Figs. $(c)$ and $(g)$ which 
have a phase of $\exp(i2p\wedge k)$
respectively. The other diagrams are divergent. For the 
$\sigma$ loop in Fig. $(b)$, we have
\begin{equation}
\begin{array}{rcl}
(b)_{\sigma}&=&\displaystyle
-\frac{3}{2}\lambda g^2v\int O_k
(k+2p_+)\cdot(k-2p_-)
D^W_kD^{\sigma}_{k+p_+}D^{\sigma}_{k-p_-}.
\end{array}
\end{equation}
Using 
\begin{equation}
\begin{array}{rcl}
O_k&=&\displaystyle c_{p\wedge(k+p_+)}\exp(ik\wedge p)
\\
&=&\displaystyle
1/2~\exp(ip_+\wedge p_-)+\cdots
\end{array}
\end{equation}
to isolate the non-$k$ oscillating part, we obtain the divergence,
\begin{equation}
\begin{array}{rcl}
(b)_{\sigma}&=&\displaystyle
-\frac{3}{4}\lambda g^2v\exp(ip_+\wedge p_-)i\deltae.
\end{array}
\end{equation}
The other two are similarly computed with the sum,
\begin{equation}
\begin{array}{rcl}
(b)&=&\displaystyle
\left[-\frac{3}{4}-\frac{1}{4}-1\right]\lambda g^2v
\exp(ip_+\wedge p_-)i\deltae.
\end{array}
\end{equation}
The same trick applies to Fig. $(d)$. Here the charge conjugated loops
contribute the same. Using obvious notations, we have
\begin{equation}
\begin{array}{rcl}
(d)&=&\displaystyle
2\cdot\left[-\frac{1}{16}-\frac{1}{16}-\frac{1}{8}\right]g^4v
\exp(ip_+\wedge p_-)i\deltae.
\end{array}
\end{equation}
Figs. $(e),(f)$ and $(h)$ are easier to compute with the result,
\begin{equation}
\begin{array}{rcl}
(e)&=&\displaystyle
+(3+1+4)\lambda^2v\exp(ip_+\wedge p_-)i\deltae,
\\
(f)&=&\displaystyle
+2\cdot 2\lambda^2v\exp(ip_+\wedge p_-)i\deltae,
\\
(h)&=&\displaystyle
+2\cdot(1/2+1/2)g^4v\exp(ip_+\wedge p_-)i\deltae.
\end{array}
\end{equation}
The divergence in the $\sigma\pi_+\pi_-$ vertex is then
\begin{equation}
\begin{array}{rcl}
iV^{\sigma\pi_+\pi_-}(p,p_+,p_-)&=&\displaystyle
i\deltae\exp(ip_+\wedge p_-)\left[
\frac{3}{2}g^4-2\lambda g^2+12\lambda^2\right]v.
\end{array}
\end{equation}
\begin{center}
\begin{picture}(400,240)(0,0)
\SetOffset(0,140)
\DashLine(80,30)(20,30){5}\DashLine(80,30)(50,70){5}
\Photon(50,70)(20,30){3}{5}
\DashLine(20,10)(20,30){5}\DashLine(80,10)(80,30){5}
\DashLine(50,90)(50,70){5}
\LongArrowArcn(50,45)(8,45,-225)
\Text(50,22)[]{$\sigma$}\Text(40,37)[]{$k$}
\Text(25,55)[]{\scriptsize$W^{\pm}$}
\Text(75,55)[]{$\pi_{\pm}$}
\Text(20,15)[r]{$\pi_{\pm}$}\Text(82,15)[l]{$\pi_{\mp}$}
\Text(52,85)[l]{$\sigma$}
\Text(22,15)[l]{$p_{\pm}$}\Text(80,15)[r]{$p_{\mp}$}
\Text(48,85)[r]{$p$}
\Text(50,0)[]{$(a)$}
\SetOffset(100,140)
\Photon(80,30)(20,30){3}{6}\DashLine(80,30)(50,70){5}
\DashLine(50,70)(20,30){5}
\DashLine(20,10)(20,30){5}\DashLine(80,10)(80,30){5}
\DashLine(50,90)(50,70){5}
\Text(50,22)[]{\scriptsize$W^-;A$}
\Text(15,60)[l]{$\sigma,\pi_0;$}\Text(15,50)[l]{$\pi_+$}
\Text(85,60)[r]{$\sigma,\pi_0;$}\Text(85,50)[r]{$\pi_+$}
\Text(20,15)[r]{$\pi_+$}\Text(82,15)[l]{$\pi_-$}
\Text(50,0)[]{$(b)$}
\SetOffset(200,140)
\DashLine(20,30)(80,30){5}\Photon(80,30)(50,70){-3}{5}
\Photon(50,70)(20,30){3}{5}
\DashLine(20,30)(20,10){5}\DashLine(80,30)(80,10){5}
\DashLine(50,70)(50,90){5}
\Text(50,22)[]{$\pi_0,\sigma$}
\Text(25,55)[]{\scriptsize$W^+$}
\Text(75,55)[]{\scriptsize$W^+$}
\Text(20,15)[r]{$\pi_+$}\Text(82,15)[l]{$\pi_-$}
\Text(50,0)[]{$(c)$}
\SetOffset(300,140)
\Photon(80,30)(20,30){3}{6}\DashLine(80,30)(50,70){5}
\Photon(50,70)(20,30){3}{5}
\DashLine(20,30)(20,10){5}\DashLine(80,30)(80,10){5}
\DashLine(50,70)(50,90){5}
\Text(50,22)[]{\scriptsize$W^{\mp};A$}
\Text(15,60)[l]{\scriptsize$Z;$}\Text(15,50)[l]{\scriptsize$W^{\pm}$}
\Text(85,60)[r]{$\sigma,\pi_0;$}\Text(85,50)[r]{$\pi_{\pm}$}
\Text(20,15)[r]{$\pi_{\pm}$}\Text(82,15)[l]{$\pi_{\mp}$}
\Text(50,0)[]{$(d)$}
\SetOffset(0,30)
\DashCArc(50,50)(20,0,360){5}
\DashLine(20,10)(50,30){5}\DashLine(50,30)(80,10){5}
\DashLine(50,70)(50,90){5}
\LongArrowArcn(50,50)(8,45,-225)\Text(32,50)[l]{$k$}
\Text(20,60)[]{$\sigma$}\Text(80,60)[]{$\sigma$}
\Text(20,50)[]{$\pi_0$}\Text(80,50)[]{$\pi_0$}
\Text(20,40)[]{$\pi_+$}\Text(80,40)[]{$\pi_+$}
\Text(20,15)[r]{$\pi_+$}\Text(80,15)[l]{$\pi_-$}
\Text(52,85)[l]{$\sigma$}
\Text(50,0)[]{$(e)$}
\SetOffset(100,30)
\DashCArc(50,50)(20,0,360){5}
\DashLine(20,10)(50,30){5}\DashLine(50,30)(80,10){5}
\DashLine(50,70)(50,90){5}
\Text(20,50)[]{$\sigma$}\Text(80,50)[]{$\pi_{\pm}$}
\Text(20,15)[r]{$\sigma$}\Text(80,15)[l]{$\pi_{\mp}$}
\Text(52,85)[l]{$\pi_{\pm}$}
\Text(50,0)[]{$(f)$}
\SetOffset(200,30)
\PhotonArc(50,50)(20,0,360){2}{12}
\DashLine(20,10)(50,30){5}\DashLine(50,30)(80,10){5}
\DashLine(50,70)(50,90){5}
\Text(20,50)[]{\scriptsize$W^+$}\Text(80,50)[]{\scriptsize$W^+$}
\Text(20,15)[r]{$\pi_+$}\Text(80,15)[l]{$\pi_-$}
\Text(52,85)[l]{$\sigma$}
\Text(50,0)[]{$(g)$}
\SetOffset(300,30)
\PhotonArc(50,50)(20,0,360){2}{12}
\DashLine(20,10)(50,30){5}\DashLine(50,30)(80,10){5}
\DashLine(50,70)(50,90){5}
\Text(20,50)[]{\scriptsize$W^{\mp}$}\Text(80,50)[]{\scriptsize$Z,A$}
\Text(20,15)[r]{$\sigma$}\Text(80,15)[l]{$\pi_{\mp}$}
\Text(52,85)[l]{$\pi_{\pm}$}
\Text(50,0)[]{$(h)$}
\SetOffset(0,0)
\Text(200,10)[]{Figure $10$. $\sigma\pi_+\pi_-$ vertex}
\end{picture}\\
\end{center}
\begin{center}
\begin{picture}(300,240)(0,0)
\SetOffset(0,140)
\DashLine(80,30)(20,30){5}\DashLine(80,30)(50,70){5}
\Photon(50,70)(20,30){3}{5}
\DashLine(20,10)(20,30){5}\DashLine(80,10)(80,30){5}
\DashLine(50,90)(50,70){5}
\LongArrowArcn(50,45)(8,45,-225)
\Text(50,22)[]{$\sigma$}\Text(40,37)[]{$k$}
\Text(25,55)[]{\scriptsize$W^{\pm}$}
\Text(75,55)[]{$\pi_{\pm}$}
\Text(20,15)[r]{$\pi_{\pm}$}\Text(82,15)[l]{$\pi_{\mp}$}
\Text(52,85)[l]{$\pi_0$}
\Text(22,15)[l]{$p_{\pm}$}\Text(80,15)[r]{$p_{\mp}$}
\Text(48,85)[r]{$p$}
\Text(50,0)[]{$(a)$}
\SetOffset(100,140)
\Photon(80,30)(20,30){3}{6}\DashLine(80,30)(50,70){5}
\DashLine(50,70)(20,30){5}
\DashLine(20,10)(20,30){5}\DashLine(80,10)(80,30){5}
\DashLine(50,90)(50,70){5}
\Text(50,22)[]{\scriptsize$W^{\mp}$}
\Text(30,55)[l]{$\sigma$}
\Text(75,55)[r]{$\pi_0$}
\Text(20,15)[r]{$\pi_{\pm}$}\Text(82,15)[l]{$\pi_{\mp}$}
\Text(50,0)[]{$(b)$}
\SetOffset(200,140)
\Photon(80,30)(20,30){3}{6}\DashLine(80,30)(50,70){5}
\Photon(50,70)(20,30){3}{5}
\DashLine(20,30)(20,10){5}\DashLine(80,30)(80,10){5}
\DashLine(50,70)(50,90){5}
\Text(50,22)[]{\scriptsize$W^{\mp};A$}
\Text(15,60)[l]{\scriptsize$Z;$}\Text(15,50)[l]{\scriptsize$W^{\pm}$}
\Text(85,60)[r]{$\sigma,\pi_0;$}\Text(85,50)[r]{$\pi_{\pm}$}
\Text(20,15)[r]{$\pi_{\pm}$}\Text(82,15)[l]{$\pi_{\mp}$}
\Text(50,0)[]{$(d)$}
\SetOffset(0,30)
\DashCArc(50,50)(20,0,360){5}
\DashLine(20,10)(50,30){5}\DashLine(50,30)(80,10){5}
\DashLine(50,70)(50,90){5}
\LongArrowArcn(50,50)(8,45,-225)\Text(32,50)[l]{$k$}
\Text(20,55)[]{$\sigma$}\Text(80,55)[]{$\pi_0$}
\Text(20,15)[r]{$\pi_+$}\Text(80,15)[l]{$\pi_-$}
\Text(52,85)[l]{$\pi_0$}
\Text(50,0)[]{$(e)$}
\SetOffset(100,30)
\DashCArc(50,50)(20,0,360){5}
\DashLine(20,10)(50,30){5}\DashLine(50,30)(80,10){5}
\DashLine(50,70)(50,90){5}
\Text(20,50)[]{$\sigma$}\Text(80,50)[]{$\pi_{\pm}$}
\Text(20,15)[r]{$\pi_0$}\Text(80,15)[l]{$\pi_{\mp}$}
\Text(52,85)[l]{$\pi_{\pm}$}
\Text(50,0)[]{$(f)$}
\SetOffset(200,30)
\PhotonArc(50,50)(20,0,360){2}{12}
\DashLine(20,10)(50,30){5}\DashLine(50,30)(80,10){5}
\DashLine(50,70)(50,90){5}
\Text(20,50)[]{\scriptsize$W^{\mp}$}\Text(80,50)[]{\scriptsize$Z,A$}
\Text(20,15)[r]{$\pi_0$}\Text(80,15)[l]{$\pi_{\mp}$}
\Text(52,85)[l]{$\pi_{\pm}$}
\Text(50,0)[]{$(h)$}
\SetOffset(0,0)
\Text(150,10)[]{Figure $11$. $\pi_0\pi_+\pi_-$ vertex}
\end{picture}\\
\end{center}

Let us compare the contributions to the $\pi_0\pi_+\pi_-$ vertex 
shown in Fig. $11$ with the above one. First, the analogs of Figs. 
$10(c)$ and $(g)$ are missing due to lack of the $\pi_0W^+W^-$
vertex. Second, $(a)$ is similarly finite. $(e)$ is also finite
because of the factor $s_{2k\wedge p}$. Third, the others always
come in a charge conjugated pair whose divergences cancel each
other. We take $(b)$ as an example. The product of the two 
$G\phi\phi$ vertices in the $W^{\mp}$ loop is
\begin{equation}
\displaystyle\mp \frac{i}{4}g^2(k+2p_+)\cdot(k-2p_-),
\end{equation}
while remaining factors are essentially the same for the 
consideration of divergence so that the divergences are cancelled
between the $W^{\mp}$ loops.

{\it 3.7 $G\phi\phi$ vertices}

We have vertices $A\pi_+\pi_-$, $Z\sigma\sigma$, $Z\sigma\pi_0$, 
$Z\pi_0\pi_0$, $W^{\pm}\pi_{\mp}\sigma$ and 
$W^{\pm}\pi_{\mp}\pi_0$ which may be divergent, and vertices
$A\sigma\sigma$, $A\sigma\pi_0$, $A\pi_0\pi_0$ and 
$Z\pi_+\pi_-$ which must be finite. For illustration we compute 
the first one of each type whose diagrams are given in Figs. 
$12$ and $13$.

Fig. $12(a)$ has a phase 
$\exp(i2k\wedge p)$ and is finite. $(d)$ is 
also finite: while the $A$ loop involves $s_{2k\wedge p}$, the $W$ 
loop is proportional to $(k+2p)_{\mu}$ and thus vanishes. 
For similar reasons $(c)$ is finite too. So we only need to 
calculate $(b)$ and $(e)$. For example,
\begin{equation}
\begin{array}{rcl}
(b)_{\pi_-}&=&\displaystyle
-\frac{i}{\sqrt{2}}g^3\int O_kP_{\mu}(k)D^W_kD^A_{k+p_+}D^A_{k-p_-},
\\
O_k&=&\displaystyle
\sin(p_+\wedge p_-+p\wedge k)\exp(ip\wedge k)
\\
&=&\displaystyle
i/2~\exp(-ip_+\wedge p_-)+\cdots,
\\
P_{\mu}(k)&=&\displaystyle
P_{\mu\alpha\beta}(p,k+p_+,p_--k)(k-p_+)^{\alpha}(k+p_-)^{\beta}
\\
&=&\displaystyle
2[k^2(p_+-p_-)_{\mu}-k_{\mu}k\cdot(p_+-p_-)]+\cdots,
\end{array}
\end{equation}
where we have ignored terms which will be finite. The divergence is
\begin{equation}
\begin{array}{rcl}
(b)_{\pi_-}&=&\displaystyle
i\deltae\frac{3}{4\sqrt{2}}g^3\exp(-ip_+\wedge p_-)(p_+-p_-)_{\mu}.
\end{array}
\end{equation}
Together with $\pi_0$ and $\sigma$ loops, we have 
\begin{equation}
\begin{array}{rcl}
(b)&=&\displaystyle
i\deltae\frac{3}{4\sqrt{2}}g^3\left[1+\frac{1}{2}+\frac{1}{2}\right]
\exp(-ip_+\wedge p_-)(p_+-p_-)_{\mu}.
\end{array}
\end{equation}
Fig. $(e)$ comes in conjugated pairs and is simpler to compute. For
example,
\begin{equation}
\begin{array}{rcl}
(e)_{\pi_-}&=&\displaystyle
+\frac{1}{\sqrt{2}}g^3\int O_k(k+2p_-)_{\mu}D^A_kD^W_{k+p_-}
\\
&=&\displaystyle
+i\deltae\frac{3}{4\sqrt{2}}g^3\exp(-ip_+\wedge p_-)p_{-\mu},
\end{array}
\end{equation}
where
\begin{equation}
\begin{array}{rcl}
O_k&=&\displaystyle
c_{p\wedge k}\exp(ip\wedge k-ip_+\wedge p_-)\\
&=&\displaystyle
1/2~\exp(-ip_+\wedge p_-)+\cdots.
\end{array}
\end{equation}
The conjugated $\pi_+$ loop is then obtained by 
$p_{-\mu}\to -p_{+\mu}$.
Including the other two pairs of loops, we have
\begin{equation}
\begin{array}{rcl}
(e)&=&\displaystyle
-i\deltae\frac{3}{4\sqrt{2}}g^3\left[1+\frac{1}{2}+\frac{1}{2}\right]
\exp(-ip_+\wedge p_-)(p_+-p_-)_{\mu},
\end{array}
\end{equation}
which cancels Fig. $(b)$ exactly so that the one loop contribution to 
the $A_{\mu}\pi_+\pi_-$ vertex is finite.
\begin{center}
\begin{picture}(300,240)(0,0)
\SetOffset(0,140)
\Photon(20,30)(80,30){3}{6}\DashLine(80,30)(50,70){5}
\DashLine(50,70)(20,30){5}
\DashLine(20,30)(20,10){5}\DashLine(80,30)(80,10){5}
\Photon(50,70)(50,90){3}{2}
\LongArrowArcn(50,45)(8,45,-225)
\Text(50,22)[]{\scriptsize$A$}\Text(40,37)[]{$k$}
\Text(25,55)[]{$\pi_+$}
\Text(75,55)[]{$\pi_+$}
\Text(20,15)[r]{$\pi_+$}\Text(82,15)[l]{$\pi_-$}
\Text(52,85)[l]{\scriptsize$A_{\mu}$}
\Text(22,15)[l]{$p_+$}\Text(80,15)[r]{$p_-$}
\Text(48,85)[r]{$p$}
\Text(50,0)[]{$(a)$}
\SetOffset(100,140)
\DashLine(20,30)(80,30){5}\Photon(80,30)(50,70){-3}{5}
\Photon(50,70)(20,30){3}{5}
\DashLine(20,30)(20,10){5}\DashLine(80,30)(80,10){5}
\Photon(50,70)(50,90){3}{2}
\Text(50,22)[]{$\pi_-;\pi_0,\sigma$}
\Text(35,55)[r]{\scriptsize$A;W^+$}
\Text(65,55)[l]{\scriptsize$A;W^+$}
\Text(20,15)[r]{$\pi_+$}\Text(82,15)[l]{$\pi_-$}
\Text(50,0)[]{$(b)$}
\SetOffset(0,30)
\DashCArc(50,50)(20,0,360){5}
\DashLine(20,10)(50,30){5}\DashLine(50,30)(80,10){5}
\Photon(50,70)(50,90){3}{2}
\LongArrowArcn(50,50)(8,45,-225)\Text(32,50)[l]{$k$}
\Text(20,55)[]{$\pi_+$}\Text(80,55)[]{$\pi_+$}
\Text(50,0)[]{$(c)$}
\SetOffset(100,30)
\PhotonArc(50,50)(20,0,360){2}{12}
\DashLine(20,10)(50,30){5}\DashLine(50,30)(80,10){5}
\Photon(50,70)(50,90){3}{2}
\Text(20,55)[]{\scriptsize$A$}\Text(80,55)[]{\scriptsize$A$}
\Text(20,45)[]{\scriptsize$W^+$}\Text(80,45)[]{\scriptsize$W^+$}
\Text(50,0)[]{$(d)$}
\SetOffset(200,30)
\PhotonArc(50,50)(20,90,270){2}{6}\DashCArc(50,50)(20,-90,90){5}
\Photon(20,10)(50,30){3}{3}\DashLine(50,30)(80,10){5}
\DashLine(50,70)(50,90){5}
\Text(20,55)[]{\scriptsize$A;$}\Text(82,55)[]{$\pi_{\mp};$}
\Text(20,45)[]{\scriptsize$W^{\pm}$}\Text(82,45)[]{$\pi_0,\sigma$}
\Text(52,85)[l]{$\pi_{\mp}$}\Text(80,15)[l]{$\pi_{\pm}$}
\Text(50,0)[]{$(e)$}
\SetOffset(0,0)
\Text(150,10)[]{Figure $12$. $A_{\mu}\pi_+\pi_-$ vertex}
\end{picture}\\
\end{center}
\begin{center}
\begin{picture}(300,240)(0,0)
\SetOffset(0,140)
\Photon(20,30)(80,30){3}{6}\DashLine(80,30)(50,70){5}
\DashLine(50,70)(20,30){5}
\DashLine(20,30)(20,10){5}\DashLine(80,30)(80,10){5}
\Photon(50,70)(50,90){3}{2}
\LongArrowArcn(50,45)(8,45,-225)
\Text(50,22)[]{\scriptsize$W^{\pm}$}\Text(40,37)[]{$k$}
\Text(25,55)[]{$\pi_{\pm}$}\Text(75,55)[]{$\pi_{\pm}$}
\Text(20,15)[r]{$\sigma$}\Text(82,15)[l]{$\sigma$}
\Text(52,85)[l]{\scriptsize$A_{\mu}$}
\Text(22,15)[l]{$p_1$}\Text(80,15)[r]{$p_2$}
\Text(48,85)[r]{$p$}
\Text(50,0)[]{$(a)$}
\SetOffset(100,140)
\DashLine(20,30)(80,30){5}\Photon(80,30)(50,70){-3}{5}
\Photon(50,70)(20,30){3}{5}
\DashLine(20,30)(20,10){5}\DashLine(80,30)(80,10){5}
\Photon(50,70)(50,90){3}{2}
\Text(50,22)[]{$\pi_{\pm}$}
\Text(35,55)[r]{\scriptsize$W^{\pm}$}
\Text(65,55)[l]{\scriptsize$W^{\pm}$}
\Text(50,0)[]{$(b)$}
\SetOffset(0,30)
\DashCArc(50,50)(20,0,360){5}
\DashLine(20,10)(50,30){5}\DashLine(50,30)(80,10){5}
\Photon(50,70)(50,90){3}{2}
\LongArrowArcn(50,50)(8,45,-225)\Text(32,50)[l]{$k$}
\Text(20,55)[]{$\pi_+$}\Text(80,55)[]{$\pi_+$}
\Text(50,0)[]{$(c)$}
\SetOffset(100,30)
\PhotonArc(50,50)(20,0,360){2}{12}
\DashLine(20,10)(50,30){5}\DashLine(50,30)(80,10){5}
\Photon(50,70)(50,90){3}{2}
\Text(20,50)[]{\scriptsize$W^+$}\Text(80,50)[]{\scriptsize$W^+$}
\Text(50,0)[]{$(d)$}
\SetOffset(200,30)
\PhotonArc(50,50)(20,90,270){2}{6}\DashCArc(50,50)(20,-90,90){5}
\Photon(20,10)(50,30){3}{3}\DashLine(50,30)(80,10){5}
\DashLine(50,70)(50,90){5}
\Text(20,50)[]{\scriptsize$W^{\pm}$}\Text(82,50)[]{$\pi_{\pm}$}
\Text(50,0)[]{$(e)$}
\SetOffset(0,0)
\Text(150,10)[]{Figure $13$. $A_{\mu}\sigma\sigma$ vertex}
\end{picture}\\
\end{center}

In contrast, the one loop contribution to the 
$A_{\mu}\sigma\sigma$ vertex is finite because each individual
diagram involves the same oscillatory phase $\exp(i2k\wedge p)$.

{\it 3.8 $GG\phi$ vertices}

We have possibly divergent vertices of
$W^{\pm}A\pi_{\mp}$, $W^{\pm}Z\pi_{\mp}$, 
$ZZ\sigma$ and $W^+W^-\sigma$ while the vertices
$AA\sigma$, $AA\pi_0$, $ZZ\pi_0$, $AZ\sigma$, $AZ\pi_0$
and $W^+W^-\pi_0$ must be finite. We show our calculation by the
last one of each type.

\begin{center}
\begin{picture}(400,240)(0,0)
\SetOffset(0,140)
\DashLine(20,30)(80,30){5}\DashLine(80,30)(50,70){5}
\DashLine(50,70)(20,30){5}
\Photon(20,30)(20,10){3}{2}\Photon(80,30)(80,10){3}{2}
\DashLine(50,70)(50,90){5}
\LongArrowArcn(50,45)(8,45,-225)
\Text(50,22)[]{$\pi_-;\sigma,\pi_0$}\Text(40,37)[]{$k$}
\Text(15,60)[l]{$\sigma,\pi_0;$}\Text(15,50)[l]{$\pi_+$}
\Text(85,60)[r]{$\sigma,\pi_0;$}\Text(85,50)[r]{$\pi_+$}
\Text(20,15)[r]{\scriptsize$W^+_{\mu}$}
\Text(82,15)[l]{\scriptsize$W^-_{\nu}$}
\Text(52,85)[l]{$\sigma$}
\Text(22,15)[l]{$p_+$}\Text(80,15)[r]{$p_-$}
\Text(48,85)[r]{$p$}
\Text(50,0)[]{$(a)$}
\SetOffset(100,140)
\DashLine(20,30)(80,30){5}\DashLine(80,30)(50,70){5}
\Photon(50,70)(20,30){3}{5}
\Photon(20,30)(20,10){3}{2}\Photon(80,30)(80,10){3}{2}
\DashLine(50,70)(50,90){5}
\Text(50,22)[]{$\pi_{\mp};\sigma$}
\Text(15,60)[l]{\scriptsize$Z;$}\Text(15,50)[l]{\scriptsize$W^{\pm}$}
\Text(85,60)[r]{$\sigma,\pi_0;$}\Text(85,50)[r]{$\pi_{\pm}$}
\Text(20,15)[r]{\scriptsize$W^{\pm}$}
\Text(82,15)[l]{\scriptsize$W^{\mp}$}
\Text(50,0)[]{$(b)$}
\SetOffset(200,140)
\Photon(20,30)(80,30){3}{5}\DashLine(80,30)(50,70){5}
\Photon(50,70)(20,30){3}{5}
\Photon(20,30)(20,10){3}{2}\Photon(80,30)(80,10){3}{2}
\DashLine(50,70)(50,90){5}
\Text(50,22)[]{\scriptsize$W^{\mp};Z,A$}
\Text(35,55)[r]{\scriptsize$Z;W^{\pm}$}
\Text(65,55)[l]{$\sigma;\pi_{\pm}$}
\Text(20,15)[r]{\scriptsize$W^{\pm}$}
\Text(82,15)[l]{\scriptsize$W^{\mp}$}
\Text(50,0)[]{$(c)$}
\SetOffset(300,140)
\Photon(20,30)(80,30){3}{5}\Photon(80,30)(50,70){-3}{5}
\Photon(50,70)(20,30){3}{5}
\Photon(20,30)(20,10){3}{2}\Photon(80,30)(80,10){3}{2}
\DashLine(50,70)(50,90){5}
\Text(50,22)[]{\scriptsize$W^-;Z,A$}
\Text(35,55)[r]{\scriptsize$Z;W^+$}
\Text(65,55)[l]{\scriptsize$Z;W^+$}
\Text(50,0)[]{$(d)$}
\SetOffset(0,30)
\DashLine(20,30)(80,30){1}\DashLine(80,30)(50,70){1}
\DashLine(50,70)(20,30){1}
\Photon(20,30)(20,10){3}{2}\Photon(80,30)(80,10){3}{2}
\DashLine(50,70)(50,90){5}
\Text(50,22)[]{$c_{\mp};c_Z,c_A$}
\Text(35,55)[r]{$c_Z;c_{\pm}$}
\Text(65,55)[l]{$c_Z;c_{\pm}$}
\Text(20,15)[r]{\scriptsize$W^{\pm}$}
\Text(82,15)[l]{\scriptsize$W^{\mp}$}
\Text(50,0)[]{$(e)$}
\SetOffset(100,30)
\DashCArc(50,50)(20,0,360){5}
\Photon(20,10)(50,30){3}{3}\Photon(50,30)(80,10){-3}{3}
\DashLine(50,70)(50,90){5}
\LongArrowArcn(50,50)(8,45,-225)\Text(32,50)[l]{$k$}
\Text(20,60)[]{$\sigma$}\Text(80,60)[]{$\sigma$}
\Text(20,50)[]{$\pi_0$}\Text(80,50)[]{$\pi_0$}
\Text(20,40)[]{$\pi_+$}\Text(80,40)[]{$\pi_+$}
\Text(50,0)[]{$(f)$}
\SetOffset(200,30)
\PhotonArc(50,50)(20,0,360){2}{12}
\Photon(20,10)(50,30){3}{3}\Photon(50,30)(80,10){-3}{3}
\DashLine(50,70)(50,90){5}
\Text(20,55)[]{\scriptsize$Z$}\Text(80,55)[]{\scriptsize$Z$}
\Text(20,45)[]{\scriptsize$W^+$}\Text(80,45)[]{\scriptsize$W^+$}
\Text(50,0)[]{$(g)$}
\SetOffset(300,30)
\PhotonArc(50,50)(20,90,270){2}{6}\DashCArc(50,50)(20,-90,90){5}
\Photon(20,10)(50,30){3}{3}\DashLine(50,30)(80,10){5}
\Photon(50,70)(50,90){3}{2}
\Text(20,55)[]{\scriptsize$W^{\pm};$}\Text(82,55)[]{$\sigma;$}
\Text(20,45)[]{\scriptsize$Z,A$}\Text(82,45)[]{$\pi_{\mp}$}
\Text(20,15)[r]{\scriptsize$W^{\pm}$}
\Text(52,85)[l]{\scriptsize$W^{\mp}$}
\Text(50,0)[]{$(h)$}
\SetOffset(0,0)
\Text(200,10)[]{Figure 14: $W^+_{\mu}W^-_{\nu}\sigma$ vertex}
\end{picture}\\
\end{center}
\begin{center}
\begin{picture}(300,240)(0,0)
\SetOffset(0,140)
\DashLine(20,30)(80,30){5}\DashLine(80,30)(50,70){5}
\DashLine(50,70)(20,30){5}
\Photon(20,30)(20,10){3}{2}\Photon(80,30)(80,10){3}{2}
\DashLine(50,70)(50,90){5}
\LongArrowArcn(50,45)(8,45,-225)
\Text(50,22)[]{$\pi_{\mp}$}\Text(40,37)[]{$k$}
\Text(25,55)[]{$\sigma$}
\Text(75,55)[]{$\pi_0$}
\Text(20,15)[r]{\scriptsize$W^{\pm}_{\mu}$}
\Text(82,15)[l]{\scriptsize$W^{\mp}_{\nu}$}
\Text(52,85)[l]{$\pi_0$}
\Text(22,15)[l]{$p_{\pm}$}\Text(80,15)[r]{$p_{\mp}$}
\Text(48,85)[r]{$p$}
\Text(50,0)[]{$(a)$}
\SetOffset(100,140)
\DashLine(20,30)(80,30){5}\DashLine(80,30)(50,70){5}
\Photon(50,70)(20,30){3}{5}
\Photon(20,30)(20,10){3}{2}\Photon(80,30)(80,10){3}{2}
\DashLine(50,70)(50,90){5}
\Text(50,22)[]{$\pi_{\mp};\sigma$}
\Text(15,60)[l]{\scriptsize$Z;$}\Text(15,50)[l]{\scriptsize$W^{\pm}$}
\Text(85,60)[r]{$\sigma,\pi_0;$}\Text(85,50)[r]{$\pi_{\pm}$}
\Text(20,15)[r]{\scriptsize$W^{\pm}$}
\Text(82,15)[l]{\scriptsize$W^{\mp}$}
\Text(50,0)[]{$(b)$}
\SetOffset(200,140)
\Photon(20,30)(80,30){3}{5}\DashLine(80,30)(50,70){5}
\Photon(50,70)(20,30){3}{5}
\Photon(20,30)(20,10){3}{2}\Photon(80,30)(80,10){3}{2}
\DashLine(50,70)(50,90){5}
\Text(50,22)[]{\scriptsize$W^{\mp};Z,A$}
\Text(35,55)[r]{\scriptsize$Z;W^{\pm}$}
\Text(65,55)[l]{$\sigma;\pi_{\pm}$}
\Text(20,15)[r]{\scriptsize$W^{\pm}$}
\Text(82,15)[l]{\scriptsize$W^{\mp}$}
\Text(50,0)[]{$(c)$}
\SetOffset(0,30)
\DashLine(20,30)(80,30){1}\DashLine(80,30)(50,70){1}
\DashLine(50,70)(20,30){1}
\Photon(20,30)(20,10){3}{2}\Photon(80,30)(80,10){3}{2}
\DashLine(50,70)(50,90){5}
\Text(50,22)[]{$c_{\mp};c_Z,c_A$}
\Text(35,55)[r]{$c_Z;c_{\pm}$}
\Text(65,55)[l]{$c_Z;c_{\pm}$}
\Text(20,15)[r]{\scriptsize$W^{\pm}$}
\Text(82,15)[l]{\scriptsize$W^{\mp}$}
\Text(50,0)[]{$(e)$}
\SetOffset(100,30)
\DashCArc(50,50)(20,0,360){5}
\Photon(20,10)(50,30){3}{3}\Photon(50,30)(80,10){-3}{3}
\DashLine(50,70)(50,90){5}
\LongArrowArcn(50,50)(8,45,-225)\Text(32,50)[l]{$k$}
\Text(20,50)[]{$\sigma$}\Text(80,50)[]{$\pi_0$}
\Text(50,0)[]{$(f)$}
\SetOffset(200,30)
\PhotonArc(50,50)(20,90,270){2}{6}\DashCArc(50,50)(20,-90,90){5}
\Photon(20,10)(50,30){3}{3}\DashLine(50,30)(80,10){5}
\Photon(50,70)(50,90){3}{2}
\Text(20,55)[]{\scriptsize$W^{\pm};$}\Text(82,55)[]{$\sigma;$}
\Text(20,45)[]{\scriptsize$A,Z$}\Text(82,45)[]{$\pi_{\mp}$}
\Text(20,15)[r]{\scriptsize$W^{\pm}$}
\Text(52,85)[l]{\scriptsize$W^{\mp}$}
\Text(50,0)[]{$(h)$}
\SetOffset(0,0)
\Text(150,10)[]{Figure 15: $W^+_{\mu}W^-_{\nu}\pi_0$ vertex}
\end{picture}\\
\end{center}

The one loop diagrams for the $W^+_{\mu}W^-_{\nu}\sigma$ vertex
are shown in Fig. $14$. Now it should be relatively easy to 
compute Fig. $(a)$, so we write down its divergence directly,
\begin{equation}
\begin{array}{rcl}
(a)&=&\displaystyle
i\deltae\lambda g^2v\gmunu\exp(ip_+\wedge p_-)\left[
\frac{3}{4}+\frac{1}{4}+0+0\right].
\end{array}
\end{equation}
The $W^{\pm}$ loops in $(b)$ are made finite by the phase 
$\exp(i2p\wedge k)$ while the divergences in the 
$Z\sigma\pi_{\mp}$ loops are cancelled by those of the
$Z\pi_0\pi_{\mp}$ loops due to a sign flip in the relevant
part of products of the $G\phi\phi$ vertices.
The $\sigma W^-$ loop of $(c)$ is,
\begin{equation}
\begin{array}{rcl}
(c)_{\sigma W^-}&=&\displaystyle
+\frac{i}{4}g^4v\int O_kP_{\mu\nu}D^W_kD^Z_{k+p_+}
D^{\sigma}_{k-p_-}
\\
&=&\displaystyle
i\deltae\frac{3}{32}g^4v\gmunu\exp(ip_+\wedge p_-)+\cdots,
\end{array}
\end{equation}
where
\begin{equation}
\begin{array}{rcl}
O_k&=&\displaystyle
\sin(p\wedge(p_--k))\exp(ik\wedge p)
\\
&=&\displaystyle
i/2\exp(ip_+\wedge p_-)+\cdots,
\\
P_{\mu\nu}&=&\displaystyle
(p-p_-+k)^{\beta}P_{\beta\mu\nu}(-k-p_+,p_+,k)
\\
&=&\displaystyle
(-k^2\gmunu+k_{\mu}k_{\nu})+\cdots.
\end{array}
\end{equation}
The remaining $\pi_{\pm}$ loops are similarly computed. Including
charge conjugated loops, we obtain,
\begin{equation}
\begin{array}{rcl}
(c)&=&\displaystyle
+i\deltae g^4v\gmunu\exp(ip_+\wedge p_-)~2\cdot\left[
\frac{3}{32}+0+\frac{3}{32}\right].
\end{array}
\end{equation}
Fig. $(d)$ is slightly more complicated. We take the $ZZW^-$ loop
as an example,
\begin{equation}
\begin{array}{rcl}
(d)_{ZZW^-}&=&\displaystyle
+\frac{1}{2}g^4v\int O_kP_{\mu\nu}D^W_kD^Z_{k+p_+}D^Z_{k-p_-}
\\
&=&\displaystyle
+i\deltae \frac{9}{8}g^4v\gmunu\exp(ip_+\wedge p_-)+\cdots,
\end{array}
\end{equation}
where
\begin{equation}
\begin{array}{rcl}
O_k&=&\displaystyle
\cos(p_+\wedge p_-+p\wedge k)\exp(ik\wedge p)
\\
&=&\displaystyle
1/2\exp(ip_+\wedge p_-)+\cdots,
\\
P_{\mu\nu}&=&\displaystyle
P_{\alpha\mu}^{~~~\beta}(-k-p_+,p_+,k)
P^{\alpha}_{~~\beta\nu}(k-p_-,-k,p_-)
\\
&=&\displaystyle
(2k^2\gmunu+10k_{\mu}k_{\nu})+\cdots.
\end{array}
\end{equation}
Including the other two contributions, we have
\begin{equation}
\begin{array}{rcl}
(d)&=&\displaystyle
+i\deltae g^4v\gmunu\exp(ip_+\wedge p_-)\left[
\frac{9}{8}+0+\frac{9}{8}\right].
\end{array}
\end{equation}
The calculation of Fig. $(e)$ is similar to $(a)$, and Figs. 
$(f)-(h)$ are the easiest of all, with the results,
\begin{equation}
\begin{array}{rcl}
(e)&=&\displaystyle
-i\deltae g^4v\gmunu\exp(ip_+\wedge p_-)~2\cdot\left[
\frac{1}{32}+0+\frac{1}{32}\right],
\\
(f)&=&\displaystyle
-i\deltae\lambda g^2v\gmunu\exp(ip_+\wedge p_-)\left[
\frac{3}{4}+\frac{1}{4}+0\right],
\\
(g)&=&\displaystyle
-i\deltae g^4v\gmunu\exp(ip_+\wedge p_-)\left[
\frac{3}{4}+\frac{3}{4}\right],
\\
(h)&=&\displaystyle
-i\deltae g^4v\gmunu\exp(ip_+\wedge p_-)~2\cdot\left[
\frac{1}{8}+0+\frac{1}{8}\right].
\end{array}
\end{equation}
In total,
\begin{equation}
\begin{array}{rcl}
iV^{W^+W^-\sigma}_{\mu\nu}(p_+,p_-,p)&=&\displaystyle
+i\deltae\frac{1}{2}g^4v\gmunu\exp(ip_+\wedge p_-).
\end{array}
\end{equation}

For the one loop $W^+_{\mu}W^-_{\nu}\pi_0$ vertex shown in
Fig. $(15)$ we briefly indicate how a finite result is achived.
The following diagrams are separately finite due to an 
oscillatory factor $\exp(i2p\wedge k)$ or $s_{2p\wedge k}$:
$W^{\pm}$ loops in $(b)$, $ZW^{\pm}\pi_{\pm}$ in $(c)$,
$c_Zc_{\pm}c_{\pm}$ in $(e)$, $Z\pi_{\mp}$ in $(h)$ and 
Fig. $(f)$. The others cancel their divergences between 
conjugated diagrams.

{\it 3.9 $GGG$ vertices}

This is the most complicated part of the calculation performed
in this section because of the involvement of $GGG$ and $GGGG$
types of vertices. The one loop contributions to the vertices 
$AAA$, $ZZZ$, $AW^+W^-$ and $ZW^+W^-$ are generally divergent
while those of $AZZ$ and $AAZ$ must be finite. Again we present
our calculation using the first one of each type as examples.

\begin{center}
\begin{picture}(300,240)(0,0)
\SetOffset(0,140)
\DashLine(20,30)(80,30){5}\DashLine(80,30)(50,70){5}
\DashLine(50,70)(20,30){5}
\Photon(20,30)(20,10){3}{2}\Photon(80,30)(80,10){3}{2}
\Photon(50,70)(50,90){3}{2}
\LongArrowArcn(50,45)(8,45,-225)
\Text(50,22)[]{$\pi_+$}\Text(40,37)[]{$k$}
\Text(25,55)[]{$\pi_+$}\Text(75,55)[]{$\pi_+$}
\Text(18,15)[r]{\scriptsize$A_{\beta}$}
\Text(82,15)[l]{\scriptsize$A_{\gamma}$}
\Text(52,85)[l]{\scriptsize$A_{\alpha}$}
\Text(22,15)[l]{$p_2$}\Text(80,15)[r]{$p_3$}
\Text(48,85)[r]{$p_1$}
\Text(50,0)[]{$(a)$}
\SetOffset(100,140)
\Photon(20,30)(80,30){3}{5}\Photon(80,30)(50,70){-3}{5}
\Photon(50,70)(20,30){3}{5}
\Photon(20,30)(20,10){3}{2}\Photon(80,30)(80,10){3}{2}
\Photon(50,70)(50,90){3}{2}
\Text(50,22)[]{\scriptsize$A,W^+$}
\Text(35,55)[r]{\scriptsize$A,W^+$}
\Text(65,55)[l]{\scriptsize$A,W^+$}
\Text(50,0)[]{$(b)$}
\SetOffset(200,140)
\DashLine(20,30)(80,30){1}\DashLine(80,30)(50,70){1}
\DashLine(50,70)(20,30){1}
\Photon(20,30)(20,10){3}{2}\Photon(80,30)(80,10){3}{2}
\Photon(50,70)(50,90){3}{2}
\Text(50,22)[]{$c_A,c_{\pm}$}
\Text(35,55)[r]{$c_A,c_{\pm}$}
\Text(65,55)[l]{$c_A,c_{\pm}$}
\Text(50,0)[]{$(c)$}
\SetOffset(0,30)
\DashCArc(50,50)(20,0,360){5}
\Photon(20,10)(50,30){3}{3}\Photon(50,30)(80,10){-3}{3}
\Photon(50,70)(50,90){3}{2}
\LongArrowArcn(50,50)(8,45,-225)\Text(32,50)[l]{$k$}
\Text(20,55)[]{$\pi_+$}\Text(80,55)[]{$\pi_+$}
\Text(20,15)[r]{\scriptsize$A_{\beta}$}
\Text(80,15)[l]{\scriptsize$A_{\gamma}$}
\Text(52,85)[l]{\scriptsize$A_{\alpha}$}
\Text(50,0)[]{$(d)$}
\SetOffset(100,30)
\PhotonArc(50,50)(20,0,360){2}{12}
\Photon(20,10)(50,30){3}{3}\Photon(50,30)(80,10){-3}{3}
\Photon(50,70)(50,90){3}{2}
\Text(20,55)[]{\scriptsize$A$}\Text(80,55)[]{\scriptsize$A$}
\Text(20,45)[]{\scriptsize$W^+$}\Text(80,45)[]{\scriptsize$W^+$}
\Text(50,0)[]{$(e)$}
\SetOffset(0,0)
\Text(150,10)[]{Figure $16$. $A_{\alpha}A_{\beta}A_{\gamma}$ vertex}
\end{picture}\\
\end{center}

Note that except for the $A$ loop in Fig. $16(b)$ there is an
additional permutated contribution for each case in $(a)-(c)$ 
and there are two additional ones for $(d)-(e)$.
We first note that $(d)$ vanishes identically. This is because
the only difference from the analog in ordinary scalar QED is
the appearance of the factor $c_{p_2\wedge p_3}$. For Fig. $(a)$
we have,
\begin{equation}
\begin{array}{rcl}
(a)^{123}&=&\displaystyle
-\frac{g^3}{2\sqrt{2}}\exp(ip_3\wedge p_2)\int 
D^W_kD^W_{k+p_2}D^W_{k-p_3}
\\
&&\displaystyle
\times(2k+p_2-p_3)_{\alpha}(2k+p_2)_{\beta}(2k-p_3)_{\gamma}
\\
&=&\displaystyle
-i\deltae\frac{g^3}{6\sqrt{2}}\exp(ip_3\wedge p_2)
P_{\alpha\beta\gamma}(p_1,p_2,p_3)+\cdots,
\end{array}
\end{equation}
Summing with the permutated one gives
\begin{equation}
\begin{array}{rcl}
(a)&=&\displaystyle(a)^{123}+(a)^{132}\\
&=&\displaystyle
-\deltae\frac{g^3}{3\sqrt{2}}\sin(p_2\wedge p_3)
P_{\alpha\beta\gamma}(p_1,p_2,p_3)+\cdots.
\end{array}
\end{equation}
The $A$ loop in $(b)$ is
\begin{equation}
\begin{array}{rcl}
(b)_A&=&\displaystyle
-i2\sqrt{2}g^3\int O_k P^t_{\alpha\beta\gamma}
D^A_kD^A_{k+p_2}D^A_{k-p_3}
\\
&=&\displaystyle
-\deltae\frac{13g^3}{4\sqrt{2}}\sin(p_1\wedge p_2)
P_{\alpha\beta\gamma}(p_1,p_2,p_3)+\cdots,
\end{array}
\end{equation}
where
\begin{equation}
\begin{array}{rcl}
O_k&=&\displaystyle
\sin(p_1\wedge(k+p_2))\sin(p_2\wedge k)\sin(k\wedge p_3)
\\
&=&\displaystyle
1/4~\sin(p_1\wedge p_2)+\cdots,
\\
P^t_{\alpha\beta\gamma}&=&\displaystyle
P^{~\rho\tau}_{\alpha}(p_1,k+p_2,p_3-k)
P_{\beta\sigma\rho}(p_2,k,-k-p_2)
\\
&&\displaystyle
\times P^{~\sigma}_{\gamma~\tau}(p_3,-k,k-p_3).
\end{array}
\end{equation}
The $W^+$ loop is similarly computed. Including its permutation,
we have the divergence
\begin{equation}
\begin{array}{rcl}
(b)_W&=&\displaystyle(b)_W^{123}+(b)_W^{132}\\
&=&\displaystyle
-i\deltae\frac{13g^3}{8\sqrt{2}}
\left[\exp(ip_3\wedge p_2)-\exp(ip_2\wedge p_3)\right]
P_{\alpha\beta\gamma}(p_1,p_2,p_3)
\\
&=&\displaystyle
-\deltae\frac{13g^3}{4\sqrt{2}}\sin(p_2\wedge p_3)
P_{\alpha\beta\gamma}(p_1,p_2,p_3).
\end{array}
\end{equation}
Fig. $(c)$ is essentially similar to $(a)$ but we must be careful with
the tensor $P_{\alpha\beta\gamma}(p_1,p_2,p_3)$. Using its properties
given in Appendix A, we have
\begin{equation}
\begin{array}{rcl}
(c)_A&=&\displaystyle(c)_A^{123}+(c)_A^{132}\\
&=&\displaystyle
-\deltae\frac{g^3}{12\sqrt{2}}\sin(p_2\wedge p_3)
\left[P_{\gamma\alpha\beta}(p_1,p_2,p_3)-
P_{\beta\alpha\gamma}(p_1,p_3,p_2)\right]
\\
&=&\displaystyle
+\deltae\frac{g^3}{12\sqrt{2}}\sin(p_2\wedge p_3)
P_{\alpha\beta\gamma}(p_1,p_2,p_3).
\end{array}
\end{equation}
The separate contribution of $c_{\pm}$ does not have
the same structure as the counterterm, 
\begin{equation}
\begin{array}{rcl}
(c)_{\pm}&=&\displaystyle(c)_{\pm}^{123}+(c)_{\pm}^{132}\\
&=&\displaystyle
\mp i\deltae\frac{g^3}{24\sqrt{2}}
\left[e^{\pm ip_3\wedge p_2}P_{\gamma\alpha\beta}(p_1,p_2,p_3)+
e^{\pm ip_2\wedge p_3}P_{\beta\alpha\gamma}(p_1,p_3,p_2)\right],
\end{array}
\end{equation}
but their sum has,
\begin{equation}
\begin{array}{rcl}
(c)&=&\displaystyle
(c)_++(c)_-\\
&=&\displaystyle
+\deltae\frac{g^3}{12\sqrt{2}}
\sin(p_2\wedge p_3)P_{\alpha\beta\gamma}(p_1,p_2,p_3).
\end{array}
\end{equation}
Now we compute the $A$ loop in Fig. $(e)$,
\begin{equation}
\begin{array}{rcl}
(e)^1_A&=&\displaystyle
-i\sqrt{2}g^3\int T_{\alpha\beta\gamma}(k,p_i)D^A_kD^A_{k+p_1},
\end{array}
\end{equation}
where
\begin{equation}
\begin{array}{rcl}
T_{\alpha\beta\gamma}(k,p_i)&=&\displaystyle
P_{\alpha}^{~\mu\nu}(p_1,k,-k-p_1)s_{p_1\wedge k}
\\
&&\displaystyle
[(g_{\mu\beta}g_{\nu\gamma}-g_{\mu\gamma}g_{\nu\beta})
s_{p_2\wedge p_3}s_{p_1\wedge k}
\\
&&\displaystyle
+(g_{\mu\gamma}g_{\nu\beta}-g_{\mu\nu}g_{\beta\gamma})
s_{p_2\wedge k}s_{p_3\wedge (k+p_1)}
\\
&&\displaystyle
+(g_{\mu\nu}g_{\beta\gamma}-g_{\mu\beta}g_{\nu\gamma})
s_{k\wedge p_3}s_{p_2\wedge (k+p_1)}].
\end{array}
\end{equation}
Isolating the non-$k$ oscillatory part as follows,
\begin{equation}
\begin{array}{rcl}
s_{p_1\wedge k}s_{p_2\wedge p_3}s_{p_1\wedge k}
&=&\displaystyle
+1/2~s_{p_2\wedge p_3}+\cdots,
\\
s_{p_1\wedge k}s_{p_2\wedge k}s_{p_3\wedge (k+p_1)}
&=&\displaystyle
-1/4~s_{p_3\wedge p_1}+\cdots,
\\
s_{p_1\wedge k}s_{k\wedge p_3}s_{p_2\wedge (k+p_1)}
&=&\displaystyle
-1/4~s_{p_1\wedge p_2}+\cdots,
\end{array}
\end{equation}
and doing loop integration by using the explicit form of 
$P_{\alpha}^{~\mu\nu}$, we arrive at
\begin{equation}
\begin{array}{rcl}
(e)^1_A&=&\displaystyle
+\deltae\frac{9}{2\sqrt{2}}g^3\sin(p_2\wedge p_3)
(p_{1\gamma}g_{\alpha\beta}-p_{1\beta}g_{\gamma\alpha})+\cdots.
\end{array}
\end{equation}
Including all permutations, we recover the correct structure,
\begin{equation}
\begin{array}{rcl}
(e)_A&=&\displaystyle
(e)^1_A+(e)^2_A+(e)^3_A\\
&=&\displaystyle
+\deltae\frac{9}{2\sqrt{2}}g^3\sin(p_2\wedge p_3)
P_{\alpha\beta\gamma}(p_1,p_2,p_3)+\cdots.
\end{array}
\end{equation}
The $W^+$ loop is simpler and turns out to have the same divergence 
as the $A$ loop. In summary, the divergence in the vertex is,
\begin{equation}
\begin{array}{rcl}
iV^{AAA}_{\alpha\beta\gamma}(p_1,p_2,p_3)&=&\displaystyle
+\deltae\frac{7}{3\sqrt{2}}g^3\sin(p_2\wedge p_3)
P_{\alpha\beta\gamma}(p_1,p_2,p_3).
\end{array}
\end{equation}

\begin{center}
\begin{picture}(400,140)(0,0)
\SetOffset(0,30)
\Photon(20,30)(80,30){3}{5}\Photon(80,30)(50,70){-3}{5}
\Photon(50,70)(20,30){3}{5}
\Photon(20,30)(20,10){3}{2}\Photon(80,30)(80,10){3}{2}
\Photon(50,70)(50,90){3}{2}
\LongArrowArcn(50,45)(8,45,-225)
\Text(50,22)[]{\scriptsize$W^+$}\Text(40,37)[]{$k$}
\Text(35,55)[r]{\scriptsize$W^+$}
\Text(65,55)[l]{\scriptsize$W^+$}
\Text(18,15)[r]{\scriptsize$Z_{\beta}$}
\Text(82,15)[l]{\scriptsize$Z_{\gamma}$}
\Text(52,85)[l]{\scriptsize$A_{\mu}$}
\Text(22,15)[l]{$p_2$}\Text(80,15)[r]{$p_3$}
\Text(48,85)[r]{$p_1$}
\Text(50,0)[]{$(b)$}
\SetOffset(100,30)
\DashLine(20,30)(80,30){1}\DashLine(80,30)(50,70){1}
\DashLine(50,70)(20,30){1}
\Photon(20,30)(20,10){3}{2}\Photon(80,30)(80,10){3}{2}
\Photon(50,70)(50,90){3}{2}
\Text(50,22)[]{$c_{\pm}$}
\Text(35,55)[r]{$c_{\pm}$}
\Text(65,55)[l]{$c_{\pm}$}
\Text(50,0)[]{$(c)$}
\SetOffset(200,30)
\PhotonArc(50,50)(20,0,360){2}{12}
\Photon(20,10)(50,30){3}{3}\Photon(50,30)(80,10){-3}{3}
\Photon(50,70)(50,90){3}{2}
\LongArrowArcn(50,50)(8,45,-225)\Text(32,50)[l]{$k$}
\Text(20,50)[]{\scriptsize$W^+$}\Text(80,50)[]{\scriptsize$W^+$}
\Text(20,15)[r]{\scriptsize$Z_{\beta}$}
\Text(80,15)[l]{\scriptsize$Z_{\gamma}$}
\Text(52,85)[l]{\scriptsize$A_{\mu}$}
\Text(50,0)[]{$(e)_A$}
\SetOffset(300,30)
\PhotonArc(50,50)(20,0,360){2}{12}
\Photon(20,10)(50,30){3}{3}\Photon(50,30)(80,10){-3}{3}
\Photon(50,70)(50,90){3}{2}
\Text(20,50)[]{\scriptsize$W^+$}\Text(80,50)[]{\scriptsize$W^+$}
\Text(20,15)[r]{\scriptsize$A_{\mu}$}
\Text(80,15)[l]{\scriptsize$Z_{\gamma}$}
\Text(52,85)[l]{\scriptsize$Z_{\alpha}$}
\Text(50,0)[]{$(e)_Z$}
\SetOffset(0,0)
\Text(150,10)[]{Figure $17$. $A_{\mu}Z_{\alpha}Z_{\beta}$ vertex}
\end{picture}\\
\end{center}

For the $A_{\mu}Z_{\alpha}Z_{\beta}$ vertex shown in Fig. $17$
we just comment that there are no analogs of Figs. $16(a)$ and $(d)$
and that the remaining diagrams are made finite by the phase
$\exp(i2k\wedge p_i)$.

{\it 3.10 $\phi c\bar{c}$ vertices}

We have the following list of possibly divergent vertices
$\sigma c_Z\bar{c}_Z$, $\sigma c_{\pm}\bar{c}_{\mp}$, 
$\pi_0 c_Z\bar{c}_Z$, $\pi_0 c_{\pm}\bar{c}_{\mp}$, 
$\pi_{\pm}c_{\mp}\bar{c}_Z$ and $\pi_{\pm}c_A\bar{c}_{\mp}$,  
and the following one which must be finite, 
$\sigma c_A\bar{c}_A$, $\sigma c_A\bar{c}_Z$, $\sigma c_Z\bar{c}_A$, 
$\pi_0 c_A\bar{c}_A$, $\pi_0 c_A\bar{c}_Z$, $\pi_0 c_Z\bar{c}_A$, 
$\pi_{\pm}c_{\mp}\bar{c}_A$ and $\pi_{\pm}c_Z\bar{c}_{\mp}$.

Let us take a sample calculation of the $\sigma c_Z\bar{c}_Z$ 
vertex shown in Fig. $18$. The $\sigma$ loop is
\begin{equation}
\begin{array}{rcl}
(\sigma)&=&\displaystyle
+\frac{1}{2}g^4v\int O_k k\cdot(k-p+\bar{q})
D^Z_kD^Z_{k-q}D^{\sigma}_{k+\bar{q}}\\
&=&\displaystyle
-i\deltae\frac{1}{8}g^4v\cos(\bar{q}\wedge q)+\cdots,
\end{array}
\end{equation}
where we have used
\begin{equation}
\begin{array}{rcl}
O_k&=&\displaystyle
\sin(p\wedge(\bar{q}+k))\sin(q\wedge k)\cos(\bar{q}\wedge k)\\
&=&\displaystyle
-1/4~\cos(\bar{q}\wedge q)+\cdots.
\end{array}
\end{equation}
The $\pi_0$ loop contributes the same, and the sum of the 
$\pi_{\pm}$ loops is
\begin{equation}
\begin{array}{rcl}
(\pi_+)+(\pi_-)&=&\displaystyle
-i\deltae\frac{1}{8}g^4v\left[\exp(i\bar{q}\wedge q)+\exp(iq\wedge\bar{q})
\right]+\cdots.
\end{array}
\end{equation}
The total divergence is then,
\begin{equation}
\begin{array}{rcl}
iV^{\sigma c_Z\bar{c}_Z}(p,q,\bar{q})&=&\displaystyle
-i\deltae\frac{1}{2}g^4v\cos(\bar{q}\wedge q).
\end{array}
\end{equation}

\begin{center}
\begin{picture}(300,140)(0,0)
\SetOffset(0,30)
\DashLine(20,30)(80,30){1}\DashLine(80,30)(50,70){5}
\Photon(50,70)(20,30){3}{5}
\DashLine(20,30)(20,10){1}\DashLine(80,30)(80,10){1}
\DashLine(50,70)(50,90){5}
\LongArrowArc(50,45)(8,-225,45)
\Text(50,22)[]{$c_Z;c_{\pm}$}\Text(40,37)[]{$k$}
\Text(35,55)[r]{\scriptsize$Z;W^{\pm}$}
\Text(65,55)[l]{$\sigma,\pi_0;\pi_{\pm}$}
\Text(18,15)[r]{$c_Z$}
\Text(82,15)[l]{$\bar{c}_Z$}
\Text(52,85)[l]{$\sigma$}
\Text(22,15)[l]{$q$}\Text(80,15)[r]{$\bar{q}$}
\Text(48,85)[r]{$p$}
\SetOffset(0,0)
\Text(50,10)[]{Figure $18$. $\sigma c_Z\bar{c}_Z$ vertex}
\SetOffset(200,30)
\DashLine(20,30)(80,30){1}\DashLine(80,30)(50,70){5}
\Photon(50,70)(20,30){3}{5}
\DashLine(20,30)(20,10){1}\DashLine(80,30)(80,10){1}
\DashLine(50,70)(50,90){5}
\LongArrowArc(50,45)(8,-225,45)
\Text(50,22)[]{$c_+$}\Text(40,37)[]{$k$}
\Text(35,55)[r]{\scriptsize$W^+$}
\Text(65,55)[l]{$\sigma,\pi_0$}
\Text(18,15)[r]{$c_Z$}
\Text(82,15)[l]{$\bar{c}_-$}
\Text(52,85)[l]{$\pi_+$}
\Text(22,15)[l]{$q$}\Text(80,15)[r]{$\bar{q}$}
\Text(48,85)[r]{$p$}
\SetOffset(150,0)
\Text(100,10)[]{Figure $19$. $\pi_+ c_Z\bar{c}_-$ vertex}
\end{picture}\\
\end{center}

As for the finite vertices, there are no apparently divergent 
diagrams at all for the vertices
$\sigma c_A\bar{c}_A$, $\sigma c_Z\bar{c}_A$, 
$\pi_0 c_A\bar{c}_A$, $\pi_0 c_Z\bar{c}_A$ and 
$\pi_{\pm}c_{\mp}\bar{c}_A$. Fig. $(19)$ is an example of the 
remaining ones which however is finite due to the appearance of
the phase $\exp(i2q\wedge k)$.

{\it 3.11 $Gc\bar{c}$ vertices}

This is the last group of vertices computed in this section. We
have generally divergent vertices 
$Ac_A\bar{c}_A$, $Ac_{\pm}\bar{c}_{\mp}$, 
$Zc_Z\bar{c}_Z$, $Zc_{\pm}\bar{c}_{\mp}$, 
$W^{\pm}c_{\mp}\bar{c}_A$, $W^{\pm}c_{\mp}\bar{c}_Z$, 
$W^{\pm}c_A\bar{c}_{\mp}$ and $W^{\pm}c_Z\bar{c}_{\mp}$, 
and finite ones
$Ac_Z\bar{c}_Z$, $Ac_A\bar{c}_Z$, $Ac_Z\bar{c}_A$, 
$Zc_A\bar{c}_A$, $Zc_A\bar{c}_Z$ and $Zc_Z\bar{c}_A$.
We illustrate our calculation by the examples shown in Figs.
$20$ and $21$.

The $c_A$ loop in Fig. $20(a)$ is
\begin{equation}
\begin{array}{rcl}
(a)_{c_A}&=&\displaystyle
-\frac{i}{\sqrt{2}}g^3\int O_kP_{\mu}D^A_kD^A_{k+\bar{q}}D^W_{k-q}
\\
&=&\displaystyle
-i\deltae \frac{3}{8\sqrt{2}}g^3\exp(i\bar{q}\wedge q)
\bar{q}_{\mu}+\cdots,
\end{array}
\end{equation}
where we have used the following
\begin{equation}
\begin{array}{rcl}
O_k&=&\displaystyle
\sin(\bar{q}\wedge k)\exp(iq\wedge k)\exp(ip\wedge (k-q))
\\
&=&\displaystyle
-i/2~\exp(i\bar{q}\wedge q)+\cdots,
\\
P_{\mu}&=&\displaystyle
P_{\alpha\mu\beta}(k+\bar{q},p,q-k)\bar{q}^{\alpha}k^{\beta}\\
&=&\displaystyle
(k^2\bar{q}_{\mu}-k\cdot\bar{q}k_{\mu})+\cdots,
\end{array}
\end{equation}
to single out the divergence. The $A$ exchange in the $c_-$ loop
is finite while the $Z$ exchange contributes the same as the $c_A$
loop. Fig. $(b)$ is simpler, so we write down the 
result directly,
\begin{equation}
\begin{array}{rcl}
(b)&=&\displaystyle
-i\deltae \frac{1}{8\sqrt{2}}g^3\exp(i\bar{q}\wedge q)
\bar{q}_{\mu}(0+1+1).
\end{array}
\end{equation}
The final result is,
\begin{equation}
\begin{array}{rcl}
iV^{W^+c_-\bar{c}_A}(p,q,\bar{q})&=&\displaystyle
-i\deltae \frac{1}{\sqrt{2}}g^3\exp(i\bar{q}\wedge q)\bar{q}_{\mu}.
\end{array}
\end{equation}
One sentence suffices for the vertex shown in Fig. $21$: all loops
are driven finite by a phase of $\exp(\pm i2k\wedge p)$ or 
$\exp(\pm i2q\wedge k)$.
\begin{center}
\begin{picture}(300,140)(0,0)
\SetOffset(0,30)
\DashLine(20,30)(80,30){1}\Photon(80,30)(50,70){-3}{5}
\Photon(50,70)(20,30){3}{5}
\DashLine(20,30)(20,10){1}\DashLine(80,30)(80,10){1}
\Photon(50,70)(50,90){3}{2}
\LongArrowArc(50,45)(8,-225,45)
\Text(50,22)[]{$c_A;c_-$}\Text(40,37)[]{$k$}
\Text(15,60)[l]{\scriptsize$W^+;$}\Text(75,60)[]{\scriptsize$A;$}
\Text(15,50)[l]{\scriptsize$A,Z$}\Text(85,50)[r]{\scriptsize$W^-$}
\Text(18,15)[r]{$c_-$}
\Text(82,15)[l]{$\bar{c}_A$}
\Text(52,85)[l]{\scriptsize$W^+_{\mu}$}
\Text(22,15)[l]{$q$}\Text(80,15)[r]{$\bar{q}$}
\Text(48,85)[r]{$p$}
\Text(50,0)[]{$(a)$}
\SetOffset(150,30)
\Photon(20,30)(80,30){3}{6}\DashLine(80,30)(50,70){1}
\DashLine(50,70)(20,30){1}
\DashLine(20,30)(20,10){1}\DashLine(80,30)(80,10){1}
\Photon(50,70)(50,90){3}{2}
\LongArrowArcn(50,45)(8,45,-225)
\Text(50,22)[]{\scriptsize$W^+;A$}\Text(40,37)[]{$k$}
\Text(10,60)[l]{$c_A,c_Z;$}\Text(85,60)[r]{$c_+;$}
\Text(15,50)[l]{$c_-$}\Text(75,50)[]{$c_A$}
\Text(50,0)[]{$(b)$}
\SetOffset(150,0)
\Text(0,10)[]{Figure $20$. $W^+_{\mu}c_-\bar{c}_A$ vertex}
\end{picture}\\
\end{center}
\begin{center}
\begin{picture}(300,140)(0,0)
\SetOffset(0,30)
\DashLine(20,30)(80,30){1}\Photon(80,30)(50,70){-3}{5}
\Photon(50,70)(20,30){3}{5}
\DashLine(20,30)(20,10){1}\DashLine(80,30)(80,10){1}
\Photon(50,70)(50,90){3}{2}
\LongArrowArc(50,45)(8,-225,45)
\Text(50,22)[]{$c_{\pm}$}\Text(40,37)[]{$k$}
\Text(35,55)[r]{\scriptsize$W^{\pm}$}
\Text(65,55)[l]{\scriptsize$W^{\pm}$}
\Text(18,15)[r]{$c_Z$}
\Text(82,15)[l]{$\bar{c}_A$}
\Text(52,85)[l]{\scriptsize$A_{\mu}$}
\Text(22,15)[l]{$q$}\Text(80,15)[r]{$\bar{q}$}
\Text(48,85)[r]{$p$}
\Text(50,0)[]{$(a)$}
\SetOffset(150,30)
\Photon(20,30)(80,30){3}{6}\DashLine(80,30)(50,70){1}
\DashLine(50,70)(20,30){1}
\DashLine(20,30)(20,10){1}\DashLine(80,30)(80,10){1}
\Photon(50,70)(50,90){3}{2}
\LongArrowArcn(50,45)(8,45,-225)
\Text(50,22)[]{\scriptsize$W^{\pm}$}\Text(40,37)[]{$k$}
\Text(35,55)[r]{$c_{\pm}$}
\Text(65,55)[l]{$c_{\pm}$}
\Text(50,0)[]{$(b)$}
\SetOffset(150,0)
\Text(0,10)[]{Figure $21$. $A_{\mu}c_Z\bar{c}_A$ vertex}
\end{picture}\\
\end{center}

{\it 3.12 Renormalization constants}

We have finished computing divergences in one loop 1PI functions in
previous subsections. Using the counterterms described in Appendix
A, we determine the renormalization constants in the MS scheme as
follows,
\begin{equation}
\begin{array}{rcl}
\delta Z_G&=&\displaystyle
\deltae\frac{19}{6}g^2,~~
\delta Z_{\phi}=\deltae 2g^2,~~
\delta Z_c=\deltae g^2,
\\
\delta Z_g&=&\displaystyle
-\deltae 2g^2,~~
\frac{\delta v}{v}=\deltae g^2,
\\
\lambda\delta Z_{\lambda}&=&\displaystyle
\deltae\left[6\lambda^2-2\lambda g^2+\frac{3}{4}g^4\right],\\
\displaystyle\lambda\frac{\delta\mu^2}{\mu^2}&=&\displaystyle
-\deltae\left[\lambda g^2+\frac{3}{8}g^4\right].
\end{array}
\end{equation}
These will be sufficient to remove all of UV divergences in Green's
functions at one loop level.
For example, the simple mass relation $m^2_Z=2m^2_W$ is preserved
by the divergent parts of their respective counterterms which 
provides a reassuring check of the consistency of the model at one
loop level.
\begin{center}
{\bf 4. Summary}
\end{center}

A potential obstacle in attemps to construct consistent models of
gauge interactions on NC spacetime is whether the renormalizability
property can be still maintained or the concept of renormalization
itself has to be modified. While there is no general proof of this 
so far, we can still get some feeling and confidence from explicit
analyses. It has been checked that the exact $U(1)$ and $U(N)$ gauge
theories and the spontaneously broken $U(1)$ gauge theory can be 
consistently renormalized at one loop order. In this work we tried 
to fill the gap by including the spontaneously broken $U(N)$ case.
We emphasize that this latter case is distinct from the former ones.
Since the gauge symmetry is partly broken we simultaneously have 
massive and massless gauge bosons which also makes the model 
closer to the
standard model of electroweak interactions. The interactions and 
masses of these particels are simply related because they are in 
the same multiplet before symmetry breaking occurs. It is not clear 
from previous studies whether these relations can still be 
accommodated at the quantum level on NC spacetime. 
This is a nontrivial problem, considering the difficulties already 
met with spontaneous breaking of global symmetries. Our explicit 
analysis shows however that this is indeed possible; just as we see 
in the usual gauge theories, with the same limited number of 
renormalization constants we can remove the UV divergences for both 
exact and spontaneously broken non-Abelian gauge theories on NC 
spacetime however complicated the latter case could be. This 
positive result supports the points of view that it is worthwhile 
to pursue further in this direction.

Although our explicit analysis is confined to the $U(2)$ case for 
both physical and technical reasons, it seems reasonable to expect
that our affirmative result also applies to the general case since 
the most important feature has already appeared in the $U(2)$ case
as discussed above, namely, the relations in interactions and 
masses among massless and massive gauge bosons as dictated by the 
group structure.

We have not discussed in this work the four point 1PI functions due
to technical reasons. A glimpse at Feynman rules makes it clear that 
the model considered here is already much more complicated than the
complete standard model of electroweak interactions. This happens in
two related ways; we have more types (almost all imaginable types)
of interactions and the interactions themselves become complicated 
due to the involvement of momentum-dependent phase factors. It is
amazing to see in the previous section how different sectors of the
model conspire to bring about a consistent result of UV divergences
so that the above complications would not spoil the renormalizability
of the model. It would not be surprising that the same magic also
occurs in four point functions since they are in a sense related to
the lower ones by the group structure and the same noncommutative
relations. We also have not included fermions. We should expect no 
problems with vector-like fermions, but it will be rather delicate 
if chiral fermions are involved due to the danger of anomalies. 
All this deserves a separate work to which we hope to return soon.

\vspace{0.5cm}
\noindent
{\bf Acknowledgements} I am grateful to H. Grosse and K. Sibold for
many helpful discussions and also to K. Sibold for carefully reading 
the manuscript.

\newpage
\begin{center}
{\bf Appendix A} Feynman rules
\end{center}

We list below the complete Feynman rules for the model. All momenta 
are incoming and shown in the parentheses of the corresponding 
particles.

Propagators (momentum $p$):
\vspace{20pt}

\begin{picture}(80,20)(0,6)
\SetOffset(25,10)
\Photon(0,0)(80,0){3}{8}
\Text(5,10)[]{$\Amu$}\Text(75,10)[]{$\Anu$}
\end{picture}\hspace{30pt}
$\displaystyle=\frac{-i}{p^2}
\left[\gmunu-(1-\xi)\frac{p_{\mu}p_{\nu}}{p^2}\right]$

\begin{picture}(80,20)(0,6)
\SetOffset(25,10)
\Photon(0,0)(80,0){3}{8}
\Text(5,10)[]{$\Zmu$}\Text(75,10)[]{$\Znu$}
\end{picture}\hspace{30pt}
$\displaystyle=\frac{-i}{p^2-m^2_Z}
\left[\gmunu-(1-\xi)\frac{p_{\mu}p_{\nu}}{p^2-\xi m^2_Z}\right]$

\begin{picture}(80,20)(0,6)
\SetOffset(25,10)
\Photon(0,0)(80,0){3}{8}
\Text(6,10)[]{$\Wpmu$}\Text(74,10)[]{$\Wmnu$}
\end{picture}\hspace{30pt}
$\displaystyle=\frac{-i}{p^2-m^2_W}
\left[\gmunu-(1-\xi)\frac{p_{\mu}p_{\nu}}{p^2-\xi m^2_W}\right]$

\begin{picture}(80,20)(0,6)
\SetOffset(25,10)
\DashLine(0,0)(80,0){5}
\Text(5,6)[]{$\sigma$}\Text(75,6)[]{$\sigma$}
\end{picture}\hspace{30pt}
$\displaystyle=\frac{i}{p^2-m^2_{\sigma}}$

\begin{picture}(80,20)(0,6)
\SetOffset(25,10)
\DashLine(0,0)(80,0){5}
\Text(5,6)[]{$\pi_0$}\Text(75,6)[]{$\pi_0$}
\end{picture}\hspace{30pt}
$\displaystyle=\frac{i}{p^2-\xi m^2_Z}$

\begin{picture}(80,20)(0,6)
\SetOffset(25,10)
\DashLine(0,0)(80,0){5}
\Text(5,6)[]{$\pi_+$}\Text(75,6)[]{$\pi_-$}
\end{picture}\hspace{30pt}
$\displaystyle=\frac{i}{p^2-\xi m^2_W}$

\begin{picture}(80,20)(0,6)
\SetOffset(25,10)
\DashArrowLine(0,0)(40,0){1}\DashArrowLine(80,0)(40,0){1}
\Text(5,6)[]{$c_A$}\Text(75,6)[]{$\bar{c}_A$}
\end{picture}\hspace{30pt}
$\displaystyle=\frac{i}{p^2}$

\begin{picture}(80,20)(0,6)
\SetOffset(25,10)
\DashArrowLine(0,0)(40,0){1}\DashArrowLine(80,0)(40,0){1}
\Text(5,6)[]{$c_Z$}\Text(75,6)[]{$\bar{c}_Z$}
\end{picture}\hspace{30pt}
$\displaystyle=\frac{i}{p^2-\xi m^2_Z}$

\begin{picture}(80,20)(0,6)
\SetOffset(25,10)
\DashArrowLine(0,0)(40,0){1}\DashArrowLine(80,0)(40,0){1}
\Text(5,6)[]{$c_{\pm}$}\Text(75,6)[]{$\bar{c}_{\mp}$}
\end{picture}\hspace{30pt}
$\displaystyle=\frac{i}{p^2-\xi m^2_W}$

$G\phi\phi$ vertices:
$$
\begin{array}{rcl}
\Amu\pi_+(p_+)\pi_-(p_-)&=&\displaystyle
\frac{ig}{\sqrt{2}}(p_+-p_-)_{\mu}\exp(-ip_+\wedge p_-) \\
\Zmu\sigma(p_1)\sigma(p_2)&=&\displaystyle
\frac{g}{\sqrt{2}}(p_1-p_2)_{\mu}\sin(p_1\wedge p_2) \\
\Zmu\pi_0(p_1)\pi_0(p_2)&=&\displaystyle
\frac{g}{\sqrt{2}}(p_1-p_2)_{\mu}\sin(p_1\wedge p_2) \\
\Zmu\sigma(p)\pi_0(p_0)&=&\displaystyle
\frac{g}{\sqrt{2}}(p-p_0)_{\mu}\cos(p\wedge p_0) \\
W^{\pm}_{\mu}\sigma(p)\pi_{\mp}(p_{\mp})&=&\displaystyle
\pm\frac{ig}{2}(p-p_{\mp})_{\mu}\exp(\mp ip\wedge p_{\mp}) \\
W^{\pm}_{\mu}\pi_0(p_0)\pi_{\mp}(p_{\mp})&=&\displaystyle
\frac{g}{2}(p_{\mp}-p_0)_{\mu}\exp(\mp ip_0\wedge p_{\mp}) \\
\end{array}
$$

$GG\phi$ vertices:
$$
\begin{array}{rcl}
W^{\pm}_{\mu}(p_{\pm})\Anu(p)\pi_{\mp}&=&\displaystyle
\frac{ig^2v}{2\sqrt{2}}\gmunu\exp(\pm ip_{\pm}\wedge p)\\
W^{\pm}_{\mu}(p_{\pm})\Znu(p)\pi_{\mp}&=&\displaystyle
\frac{ig^2v}{2\sqrt{2}}\gmunu\exp(\mp ip_{\pm}\wedge p)\\
\Wpmu(p_+)\Wmnu(p_-)\sigma&=&\displaystyle
\frac{ig^2v}{2}\gmunu\exp(ip_+\wedge p_-)\\
\Zmu(p_1)\Znu(p_2)\sigma&=&\displaystyle
ig^2v\gmunu\cos(p_1\wedge p_2)\\
\end{array}
$$

$GG\phi\phi$ vertices:
$$
\begin{array}{rcl}
\Amu(k_1)\Anu(k_2)\pi_+(p_+)\pi_-(p_-)&=&\displaystyle
ig^2\gmunu\cos(k_1\wedge k_2)\exp(-ip_+\wedge p_-)\\
\Zmu(k_1)\Znu(k_2)\sigma(p_1)\sigma(p_2)&=&\displaystyle
ig^2\gmunu\cos(k_1\wedge k_2)\cos(p_1\wedge p_2)\\
\Zmu(k_1)\Znu(k_2)\pi_0(p_1)\pi_0(p_2)&=&\displaystyle
ig^2\gmunu\cos(k_1\wedge k_2)\cos(p_1\wedge p_2)\\
\Zmu(k_1)\Znu(k_2)\sigma(p)\pi_0(p_0)&=&\displaystyle
-ig^2\gmunu\cos(k_1\wedge k_2)\sin(p\wedge p_0)\\
\Wpmu(k_+)\Wmnu(k_-)\pi_+(p_+)\pi_-(p_-)&=&\displaystyle
\frac{ig^2}{2}\gmunu\exp(-ik_+\wedge k_-)\exp(-ip_+\wedge p_-)\\
\Wpmu(k_+)\Wmnu(k_-)\sigma(p_1)\sigma(p_2)&=&\displaystyle
\frac{ig^2}{2}\gmunu\exp(ik_+\wedge k_-)\cos(p_1\wedge p_2)\\
\Wpmu(k_+)\Wmnu(k_-)\pi_0(p_1)\pi_0(p_2)&=&\displaystyle
\frac{ig^2}{2}\gmunu\exp(ik_+\wedge k_-)\cos(p_1\wedge p_2)\\
\Wpmu(k_+)\Wmnu(k_-)\sigma(p)\pi_0(p_0)&=&\displaystyle
-\frac{ig^2}{2}\gmunu\exp(ik_+\wedge k_-)\sin(p\wedge p_0)\\
W^{\pm}(k_{\pm})\Anu(k)\sigma(p)\pi_{\mp}(p_{\mp})&=&\displaystyle
\frac{ig^2}{2\sqrt{2}}\gmunu\exp(\pm ik_{\pm}\wedge k)
\exp(\pm ip_{\mp}\wedge p)\\
W^{\pm}(k_{\pm})\Znu(k)\sigma(p)\pi_{\mp}(p_{\mp})&=&\displaystyle
\frac{ig^2}{2\sqrt{2}}\gmunu\exp(\mp ik_{\pm}\wedge k)
\exp(\pm ip_{\mp}\wedge p)\\
W^{\pm}(k_{\pm})\Anu(k)\pi_0(p)\pi_{\mp}(p_{\mp})&=&\displaystyle
\mp\frac{g^2}{2\sqrt{2}}\gmunu\exp(\pm ik_{\pm}\wedge k)
\exp(\pm ip_{\mp}\wedge p)\\
W^{\pm}(k_{\pm})\Znu(k)\pi_0(p)\pi_{\mp}(p_{\mp})&=&\displaystyle
\mp\frac{g^2}{2\sqrt{2}}\gmunu\exp(\mp ik_{\pm}\wedge k)
\exp(\pm ip_{\mp}\wedge p)\\
\end{array}
$$

$GGG$ vertices:
$$
\begin{array}{rcl}
\Aalf(k_1)\Abet(k_2)\Agam(k_3) &=&\displaystyle
-\sqrt{2}g\sin(k_1\wedge k_2)P_{\alpha\beta\gamma}(k_1,k_2,k_3)\\
\Zalf(k_1)\Zbet(k_2)\Zgam(k_3) &=&\displaystyle
-\sqrt{2}g\sin(k_1\wedge k_2)P_{\alpha\beta\gamma}(k_1,k_2,k_3)\\
\Amu(k)W^+_{\alpha}(k_+)W^-_{\beta}(k_-) &=&\displaystyle
-\frac{ig}{\sqrt{2}}\exp(-ik_+\wedge k_-)P_{\mu\alpha\beta}(k,k_+,k_-)\\
\Zmu(k)W^+_{\alpha}(k_+)W^-_{\beta}(k_-) &=&\displaystyle
+\frac{ig}{\sqrt{2}}\exp(+ik_+\wedge k_-)P_{\mu\alpha\beta}(k,k_+,k_-)\\
\end{array}
$$
where
$$
P_{\alpha\beta\gamma}(k_1,k_2,k_3)=(k_1-k_2)_{\gamma}g_{\alpha\beta}+
(k_2-k_3)_{\alpha}g_{\beta\gamma}+(k_3-k_1)_{\beta}g_{\gamma\alpha}.
$$
Some simple properties of it are useful:
$$
\begin{array}{l}
P_{\alpha\beta\gamma}(k_1,k_2,k_3)
=-P_{\alpha\gamma\beta}(k_1,k_3,k_2)
=-P_{\beta\alpha\gamma}(k_2,k_1,k_3)
=-P_{\gamma\beta\alpha}(k_3,k_2,k_1),
\end{array}
$$
$$
P_{\alpha\beta\gamma}(k_1,k_2,k_3)+P_{\gamma\alpha\beta}(k_1,k_2,k_3)
=P_{\beta\alpha\gamma}(k_1,k_3,k_2).
$$

$GGGG$ vertices:
$$
\begin{array}{l}
\Aalf(k_1)\Abet(k_2)\Amu(k_3)\Anu(k_4)=
\Zalf(k_1)\Zbet(k_2)\Zmu(k_3)\Znu(k_4)\\
=\displaystyle
-i2g^2\{(g_{\mu\alpha}g_{\nu\beta}-g_{\mu\beta}g_{\nu\alpha})
\sin(k_1\wedge k_2)\sin(k_3\wedge k_4)\\
\displaystyle
~~~~~~~~+(g_{\mu\beta}g_{\nu\alpha}-g_{\mu\nu}g_{\alpha\beta})
\sin(k_3\wedge k_1)\sin(k_2\wedge k_4)\\
\displaystyle
~~~~~~~~+(g_{\mu\nu}g_{\alpha\beta}-g_{\mu\alpha}g_{\nu\beta})
\sin(k_1\wedge k_4)\sin(k_2\wedge k_3)\}\\
\\
W^-_{\alpha}(k_1)W^-_{\beta}(k_2)W^+_{\mu}(k_3)W^+_{\nu}(k_4)\\
=\displaystyle
ig^2(2\gmunu g_{\alpha\beta}-g_{\mu\alpha}g_{\nu\beta}
-g_{\mu\beta}g_{\nu\alpha})\cos(k_1\wedge k_3+k_2\wedge k_4)\\
\\
W^+_{\alpha}(k_+)W^-_{\beta}(k_-)\Amu(k_1)\Anu(k_2)\\
=\displaystyle
\frac{ig^2}{2}\{(-2\gmunu g_{\alpha\beta}+g_{\mu\alpha}g_{\nu\beta}
+g_{\mu\beta}g_{\nu\alpha})\cos(k_1\wedge k_2)\\
~~~~~~~~~+(g_{\mu\beta}g_{\nu\alpha}-g_{\mu\alpha}g_{\nu\beta})3i
\sin(k_1\wedge k_2)\}\exp(-ik_+\wedge k_-)\\
\\
W^+_{\alpha}(k_+)W^-_{\beta}(k_-)\Zmu(k_1)\Znu(k_2)\\
=\displaystyle
\frac{ig^2}{2}\{(-2\gmunu g_{\alpha\beta}+g_{\mu\alpha}g_{\nu\beta}
+g_{\mu\beta}g_{\nu\alpha})\cos(k_1\wedge k_2)\\
~~~~~~~~~+(g_{\mu\alpha}g_{\nu\beta}-g_{\mu\beta}g_{\nu\alpha})3i
\sin(k_1\wedge k_2)\}\exp(ik_+\wedge k_-)\\
\\
W^+_{\alpha}(k_+)W^-_{\beta}(k_-)\Zmu(k_1)\Anu(k_2)\\
=\displaystyle
\frac{ig^2}{2}(2\gmunu g_{\alpha\beta}-g_{\mu\alpha}g_{\nu\beta}
-g_{\mu\beta}g_{\nu\alpha})\exp(ik_+\wedge k_2)\exp(ik_-\wedge k_1)\\
\end{array}
$$

$\phi\phi\phi$ vertices:
$$
\begin{array}{rcl}
\sigma\sigma(p_1)\sigma(p_2)&=&\displaystyle
-i6\lambda v\cos(p_1\wedge p_2)\\
\sigma\pi_0(p_1)\pi_0(p_2)&=&\displaystyle
-i2\lambda v\cos(p_1\wedge p_2)\\
\sigma\pi_+(p_+)\pi_-(p_-)&=&\displaystyle
-i2\lambda v\exp(ip_+\wedge p_-)\\
\end{array}
$$

$\phi\phi\phi\phi$ vertices:
$$
\begin{array}{l}
\sigma(p_1)\sigma(p_2)\sigma(p_3)\sigma(p_4)=
\pi_0(p_1)\pi_0(p_2)\pi_0(p_3)\pi_0(p_4)\\
=\displaystyle
-i2\lambda[\cos(p_1\wedge p_2)\cos(p_3\wedge p_4)
+\cos(p_1\wedge p_3)\cos(p_2\wedge p_4)\\
\displaystyle
~~~~~~+\cos(p_1\wedge p_4)\cos(p_2\wedge p_3)]\\
\\
\sigma(p_1)\sigma(p_2)\pi_0(k_1)\pi_0(k_2)\\
\displaystyle
=-i2\lambda\{2\cos(p_1\wedge p_2)\cos(k_1\wedge k_2)
-\cos(p_1\wedge k_1+p_2\wedge k_2)\}\\
\\
\pi_+(p_1)\pi_+(p_2)\pi_-(k_1)\pi_-(k_2)
\displaystyle
=-i4\lambda\cos(p_1\wedge k_1+p_2\wedge k_2)\\
\\
\sigma(p_1)\sigma(p_2)\pi_+(k_+)\pi_-(k_-)=
\pi_0(p_1)\pi_0(p_2)\pi_+(k_+)\pi_-(k_-)\\
\displaystyle=-i2\lambda\cos(p_1\wedge p_2)\exp(ik_+\wedge k_-)\\
\\
\sigma(p)\pi_0(k_0)\pi_+(k_+)\pi_-(k_-)
\displaystyle=-i2\lambda\sin(p\wedge k_0)\exp(ik_+\wedge k_-)
\end{array}
$$

$\phi c\bar{c}$ vertices:
$$
\begin{array}{rcl}
\sigma c_Z(p)\bar{c}_Z(\bar{p})&=&
\displaystyle -i\xi g^2v/2~\cos(\bar{p}\wedge p)\\
\sigma c_{\pm}(p)\bar{c}_{\mp}(\bar{p})&=&
\displaystyle -i\xi g^2v/4~\exp(\mp i\bar{p}\wedge p)\\
\pi_0 c_Z(p)\bar{c}_Z(\bar{p})&=&
\displaystyle -i\xi g^2v/2~\sin(\bar{p}\wedge p)\\
\pi_0 c_{\pm}(p)\bar{c}_{\mp}(\bar{p})&=&
\displaystyle \pm\xi g^2v/4~\exp(\mp i\bar{p}\wedge p)\\
\pi_{\pm} c_{\mp}(p)\bar{c}_Z(\bar{p})&=&
\displaystyle -i\xi g^2v\sqrt{2}/4~\exp(\mp i\bar{p}\wedge p)\\
\pi_{\pm} c_A(p)\bar{c}_{\mp}(\bar{p})&=&
\displaystyle -i\xi g^2v\sqrt{2}/4~\exp(\mp i\bar{p}\wedge p)\\
\end{array}
$$

$Gc\bar{c}$ vertices:
$$
\begin{array}{rcl}
\Amu c_A(p)\bar{c}_A(\bar{p})&=&
\displaystyle \sqrt{2}g\bar{p}_{\mu}\sin(\bar{p}\wedge p)\\
\Amu c_{\pm}(p)\bar{c}_{\mp}(\bar{p})&=&
\displaystyle \mp ig/\sqrt{2}~\bar{p}_{\mu}\exp(\pm i\bar{p}\wedge p)\\
\Zmu c_Z(p)\bar{c}_Z(\bar{p})&=&
\displaystyle \sqrt{2}g\bar{p}_{\mu}\sin(\bar{p}\wedge p)\\
\Zmu c_{\pm}(p)\bar{c}_{\mp}(\bar{p})&=&
\displaystyle \pm ig/\sqrt{2}~\bar{p}_{\mu}\exp(\mp i\bar{p}\wedge p)\\
W^{\pm}_{\mu}c_{\mp}(p)\bar{c}_A(\bar{p})&=&
\displaystyle \mp ig/\sqrt{2}~\bar{p}_{\mu}\exp(\pm i\bar{p}\wedge p)\\
W^{\pm}_{\mu}c_A(p)\bar{c}_{\mp}(\bar{p})&=&
\displaystyle \pm ig/\sqrt{2}~\bar{p}_{\mu}\exp(\mp i\bar{p}\wedge p)\\
W^{\pm}_{\mu}c_{\mp}(p)\bar{c}_Z(\bar{p})&=&
\displaystyle \pm ig/\sqrt{2}~\bar{p}_{\mu}\exp(\mp i\bar{p}\wedge p)\\
W^{\pm}_{\mu}c_Z(p)\bar{c}_{\mp}(\bar{p})&=&
\displaystyle \mp ig/\sqrt{2}~\bar{p}_{\mu}\exp(\pm i\bar{p}\wedge p)\\
\end{array}
$$

Counterterms for self-energies and mixings are listed below. Note the
momentum $p$ is the incoming momentum of the gauge boson in the
$G\phi$ mixing.

\begin{picture}(80,20)(0,6)
\SetOffset(25,10)
\DashLine(0,0)(80,0){5}
\Text(5,6)[]{$\sigma$}\Text(80,0)[]{$\times$}
\end{picture}\hspace{30pt}
$\displaystyle=i\lambda v^3
\left[\frac{\delta\mu^2}{\mu^2}-\delta Z_{\lambda}-\frac{2\delta v}{v}\right]$

\begin{picture}(80,20)(0,6)
\SetOffset(25,10)
\DashLine(0,0)(80,0){5}
\Text(5,6)[]{$\sigma$}\Text(80,6)[]{$\sigma$}\Text(40,0)[]{$\times$}
\end{picture}\hspace{30pt}
$\displaystyle=ip^2\delta Z_{\phi}
-im^2_{\sigma}\left[-\frac{\delta\mu^2}{2\mu^2}
+\frac{3}{2}\delta Z_{\lambda}+\frac{3\delta v}{v}\right]$

\begin{picture}(80,20)(0,6)
\SetOffset(25,10)
\DashLine(0,0)(80,0){5}
\Text(5,6)[]{$\pi_0$}\Text(80,6)[]{$\pi_0$}\Text(40,0)[]{$\times$}
\end{picture}\hspace{30pt}
$\displaystyle=ip^2\delta Z_{\phi}
-im^2_{\sigma}\left[-\frac{\delta\mu^2}{2\mu^2}
+\frac{1}{2}\delta Z_{\lambda}+\frac{\delta v}{v}\right]$

\begin{picture}(80,20)(0,6)
\SetOffset(25,10)
\DashLine(0,0)(80,0){5}
\Text(5,6)[]{$\pi_+$}\Text(80,6)[]{$\pi_-$}\Text(40,0)[]{$\times$}
\end{picture}\hspace{30pt}
$\displaystyle=ip^2\delta Z_{\phi}
-im^2_{\sigma}\left[-\frac{\delta\mu^2}{2\mu^2}
+\frac{1}{2}\delta Z_{\lambda}+\frac{\delta v}{v}\right]$

\begin{picture}(80,20)(0,6)
\SetOffset(25,10)
\Photon(0,0)(80,0){3}{8}
\Text(5,10)[]{$\Amu$}\Text(80,10)[]{$\Anu$}\Text(40,0)[]{$\times$}
\end{picture}\hspace{30pt}
$\displaystyle=i(p_{\mu}p_{\nu}-p^2\gmunu)\delta Z_G$

\begin{picture}(80,20)(0,6)
\SetOffset(25,10)
\Photon(0,0)(80,0){3}{8}
\Text(5,10)[]{$\Zmu$}\Text(80,10)[]{$\Znu$}\Text(40,0)[]{$\times$}
\end{picture}\hspace{30pt}
$\displaystyle=i(p_{\mu}p_{\nu}-p^2\gmunu)\delta Z_G
+i\gmunu m^2_Z\left[2\delta Z_g+\frac{2\delta v}{v}+\delta Z_{\phi}\right]$

\begin{picture}(80,20)(0,6)
\SetOffset(25,10)
\Photon(0,0)(80,0){3}{8}
\Text(5,10)[]{$\Wpmu$}\Text(80,10)[]{$\Wmnu$}\Text(40,0)[]{$\times$}
\end{picture}\hspace{30pt}
$\displaystyle=i(p_{\mu}p_{\nu}-p^2\gmunu)\delta Z_G
+i\gmunu m^2_W\left[2\delta Z_g+\frac{2\delta v}{v}+\delta Z_{\phi}\right]$

\begin{picture}(80,20)(0,6)
\SetOffset(25,10)
\Photon(0,0)(40,0){3}{4}\DashLine(40,0)(80,0){5}
\Text(5,10)[]{$\Zmu$}\Text(80,10)[]{$\pi_0$}\Text(40,0)[]{$\times$}
\end{picture}\hspace{30pt}
$\displaystyle=m_Zp_{\mu}\left[\delta Z_g+\frac{\delta v}{v}+
\delta Z_{\phi}\right]$

\begin{picture}(80,20)(0,6)
\SetOffset(25,10)
\Photon(0,0)(40,0){3}{4}\DashLine(40,0)(80,0){5}
\Text(5,10)[]{$W^{\pm}_{\mu}$}\Text(80,10)[]{$\pi_{\mp}$}
\Text(40,0)[]{$\times$}
\end{picture}\hspace{30pt}
$\displaystyle=\pm im_Wp_{\mu}
\left[\delta Z_g+\frac{\delta v}{v}+\delta Z_{\phi}\right]$

\begin{picture}(80,20)(0,6)
\SetOffset(25,10)
\DashArrowLine(0,0)(40,0){1}\DashArrowLine(80,0)(40,0){1}
\Text(5,6)[]{$c_A$}\Text(75,6)[]{$\bar{c}_A$}
\Text(40,0)[]{$\times$}
\end{picture}\hspace{30pt}
$\displaystyle=ip^2\delta Z_c$

\begin{picture}(80,20)(0,6)
\SetOffset(25,10)
\DashArrowLine(0,0)(40,0){1}\DashArrowLine(80,0)(40,0){1}
\Text(5,6)[]{$c_Z$}\Text(75,6)[]{$\bar{c}_Z$}
\Text(40,0)[]{$\times$}
\end{picture}\hspace{30pt}
$\displaystyle=ip^2\delta Z_c-i\xi m^2_Z
\left[\delta Z_g+\frac{\delta v}{v}+\delta Z_c\right]$

\begin{picture}(80,20)(0,6)
\SetOffset(25,10)
\DashArrowLine(0,0)(40,0){1}\DashArrowLine(80,0)(40,0){1}
\Text(5,6)[]{$c_{\pm}$}\Text(75,6)[]{$\bar{c}_{\mp}$}
\Text(40,0)[]{$\times$}
\end{picture}\hspace{30pt}
$\displaystyle=ip^2\delta Z_c-i\xi m^2_W
\left[\delta Z_g+\frac{\delta v}{v}+\delta Z_c\right]$

The counterterms for vertices are obtained by attaching the 
following factors to the corresponding Feynman rules.
$$
\begin{array}{rl}
G\phi\phi:&(Z_{\phi}Z_g-1)=\delta Z_{\phi}+\delta Z_g, \\
GG\phi:&[Z_{\phi}Z_g^2(1+\delta v/v)-1]=
\delta Z_{\phi}+2\delta Z_g+\delta v/v, \\
GG\phi\phi:&(Z_{\phi}Z_g^2-1)=\delta Z_{\phi}+2\delta Z_g, \\
GGG:&(Z_GZ_g-1)=\delta Z_G+\delta Z_g, \\
GGGG:&(Z_GZ_g^2-1)=\delta Z_G+2\delta Z_g, \\
\phi\phi\phi:&[Z_{\lambda}(1+\delta v/v)-1]=
\delta Z_{\lambda}+\delta v/v, \\
\phi\phi\phi\phi:&(Z_{\lambda}-1)=\delta Z_{\lambda}, \\
\phi c\bar{c}:&(Z_cZ_g-1)=\delta Z_c+\delta Z_g,\\
G c\bar{c}:&(Z_cZ_g-1)=\delta Z_c+\delta Z_g.
\end{array}
$$

\begin{center}
{\bf Appendix B} One loop diagrams for 1PI functions
\end{center}

We show below topologically different diagrams in which the wavy, dashed
and dotted lines represent the gauge, scalar and ghost fields respectively.
For a concrete vertex all possible assignments of fields must be
included. Diagrams with an `` f '' are finite by power counting.

$\phi$ self-energy:

\begin{center}
\begin{picture}(400,120)(0,0)
\SetOffset(0,60)
\DashCArc(50,50)(20,0,360){5}
\DashLine(10,50)(30,50){5}\DashLine(70,50)(90,50){5}
\SetOffset(100,60)
\PhotonArc(50,50)(20,0,180){2}{6}\DashCArc(50,50)(20,180,360){5}
\DashLine(10,50)(30,50){5}\DashLine(70,50)(90,50){5}
\SetOffset(200,60)
\PhotonArc(50,50)(20,0,360){2}{12}
\DashLine(10,50)(30,50){5}\DashLine(70,50)(90,50){5}
\SetOffset(300,60)
\DashCArc(50,50)(20,0,360){1}
\DashLine(10,50)(30,50){5}\DashLine(70,50)(90,50){5}

\SetOffset(0,0)
\DashCArc(50,60)(20,0,360){5}
\DashLine(10,40)(90,40){5}
\SetOffset(100,0)
\PhotonArc(50,62)(20,-90,270){2}{12}
\DashLine(10,40)(90,40){5}
\end{picture}\\
\end{center}

$G$ self-energy:

\begin{center}
\begin{picture}(400,120)(0,0)
\SetOffset(0,60)
\DashCArc(50,50)(20,0,360){5}
\Photon(10,50)(30,50){3}{3}\Photon(70,50)(90,50){3}{3}
\SetOffset(100,60)
\PhotonArc(50,50)(20,0,180){2}{6}\DashCArc(50,50)(20,180,360){5}
\Photon(10,50)(30,50){3}{3}\Photon(70,50)(90,50){3}{3}
\SetOffset(200,60)
\PhotonArc(50,50)(20,0,360){2}{12}
\Photon(10,50)(30,50){3}{3}\Photon(70,50)(90,50){3}{3}
\SetOffset(300,60)
\DashCArc(50,50)(20,0,360){1}
\Photon(10,50)(30,50){3}{3}\Photon(70,50)(90,50){3}{3}

\SetOffset(0,0)
\DashCArc(50,62)(20,0,360){5}
\Photon(10,40)(90,40){3}{10}
\SetOffset(100,0)
\PhotonArc(50,64)(20,-90,270){2}{12}
\Photon(10,40)(90,40){3}{10}
\end{picture}\\
\end{center}

$G\phi$ mixing:

\begin{center}
\begin{picture}(400,60)(0,0)
\DashCArc(50,50)(20,0,360){5}
\Photon(10,50)(30,50){3}{3}\DashLine(70,50)(90,50){5}
\SetOffset(100,0)
\PhotonArc(50,50)(20,0,180){2}{6}\DashCArc(50,50)(20,180,360){5}
\Photon(10,50)(30,50){3}{3}\DashLine(70,50)(90,50){5}
\SetOffset(200,0)
\PhotonArc(50,50)(20,0,360){2}{12}
\Photon(10,50)(30,50){3}{3}\DashLine(70,50)(90,50){5}
\SetOffset(300,0)
\DashCArc(50,50)(20,0,360){1}
\Photon(10,50)(30,50){3}{3}\DashLine(70,50)(90,50){5}
\end{picture}\\
\end{center}

$c$ self-energy:

\begin{center}
\begin{picture}(400,60)(0,0)
\DashCArc(50,40)(20,0,180){5}
\DashLine(10,40)(90,40){1}
\SetOffset(100,0)
\PhotonArc(50,40)(20,0,180){2}{6}
\DashLine(10,40)(90,40){1}
\end{picture}\\
\end{center}

$\phi\phi\phi$ vertex:

\begin{center}
\begin{picture}(400,180)(0,0)
\SetOffset(0,90)
\DashLine(20,30)(80,30){5}\DashLine(80,30)(50,70){5}
\DashLine(50,70)(20,30){5}
\DashLine(20,30)(20,10){5}\DashLine(80,30)(80,10){5}
\DashLine(50,70)(50,90){5}
\Text(50,50)[]{f}
\SetOffset(80,90)
\DashLine(20,30)(80,30){5}\DashLine(80,30)(50,70){5}
\Photon(50,70)(20,30){3}{5}
\DashLine(20,30)(20,10){5}\DashLine(80,30)(80,10){5}
\DashLine(50,70)(50,90){5}
\SetOffset(160,90)
\DashLine(20,30)(80,30){5}\Photon(80,30)(50,70){-3}{5}
\Photon(50,70)(20,30){3}{5}
\DashLine(20,30)(20,10){5}\DashLine(80,30)(80,10){5}
\DashLine(50,70)(50,90){5}
\SetOffset(240,90)
\Photon(20,30)(80,30){3}{5}\Photon(80,30)(50,70){-3}{5}
\Photon(50,70)(20,30){3}{5}
\DashLine(20,30)(20,10){5}\DashLine(80,30)(80,10){5}
\DashLine(50,70)(50,90){5}
\Text(50,50)[]{f}
\SetOffset(320,90)
\DashLine(20,30)(80,30){1}\DashLine(80,30)(50,70){1}
\DashLine(50,70)(20,30){1}
\DashLine(20,30)(20,10){5}\DashLine(80,30)(80,10){5}
\DashLine(50,70)(50,90){5}
\Text(50,50)[]{f}

\SetOffset(0,0)
\DashCArc(50,50)(20,0,360){5}
\DashLine(20,10)(50,30){5}\DashLine(50,30)(80,10){5}
\DashLine(50,70)(50,90){5}
\SetOffset(80,0)
\PhotonArc(50,50)(20,0,360){2}{12}
\DashLine(20,10)(50,30){5}\DashLine(50,30)(80,10){5}
\DashLine(50,70)(50,90){5}
\end{picture}\\
\end{center}

$G\phi\phi$ vertex:

\begin{center}
\begin{picture}(400,180)(0,0)
\SetOffset(0,90)
\DashLine(20,30)(80,30){5}\DashLine(80,30)(50,70){5}
\DashLine(50,70)(20,30){5}
\DashLine(20,30)(20,10){5}\DashLine(80,30)(80,10){5}
\Photon(50,70)(50,90){3}{2}
\Text(50,50)[]{f}
\SetOffset(80,90)
\DashLine(20,30)(80,30){5}\DashLine(80,30)(50,70){5}
\Photon(50,70)(20,30){3}{5}
\DashLine(20,30)(20,10){5}\DashLine(80,30)(80,10){5}
\Photon(50,70)(50,90){3}{2}
\Text(50,50)[]{f}
\SetOffset(160,90)
\Photon(20,30)(80,30){3}{5}\DashLine(80,30)(50,70){5}
\DashLine(50,70)(20,30){5}
\DashLine(20,30)(20,10){5}\DashLine(80,30)(80,10){5}
\Photon(50,70)(50,90){3}{2}
\SetOffset(240,90)
\Photon(20,30)(80,30){3}{5}\DashLine(80,30)(50,70){5}
\Photon(50,70)(20,30){3}{5}
\DashLine(20,30)(20,10){5}\DashLine(80,30)(80,10){5}
\Photon(50,70)(50,90){3}{2}
\Text(50,50)[]{f}
\SetOffset(320,90)
\DashLine(20,30)(80,30){5}\Photon(80,30)(50,70){-3}{5}
\Photon(50,70)(20,30){3}{5}
\DashLine(20,30)(20,10){5}\DashLine(80,30)(80,10){5}
\Photon(50,70)(50,90){3}{2}

\SetOffset(0,0)
\Photon(20,30)(80,30){3}{5}\Photon(80,30)(50,70){-3}{5}
\Photon(50,70)(20,30){3}{5}
\DashLine(20,30)(20,10){5}\DashLine(80,30)(80,10){5}
\Photon(50,70)(50,90){3}{2}
\Text(50,50)[]{f}
\SetOffset(80,0)
\DashLine(20,30)(80,30){1}\DashLine(80,30)(50,70){1}
\DashLine(50,70)(20,30){1}
\DashLine(20,30)(20,10){5}\DashLine(80,30)(80,10){5}
\Photon(50,70)(50,90){3}{2}
\Text(50,50)[]{f}
\SetOffset(160,0)
\DashCArc(50,50)(20,0,360){5}
\DashLine(20,10)(50,30){5}\DashLine(50,30)(80,10){5}
\Photon(50,70)(50,90){3}{2}
\SetOffset(240,0)
\PhotonArc(50,50)(20,0,360){2}{12}
\DashLine(20,10)(50,30){5}\DashLine(50,30)(80,10){5}
\Photon(50,70)(50,90){3}{2}
\SetOffset(320,0)
\PhotonArc(50,50)(20,90,270){2}{6}\DashCArc(50,50)(20,-90,90){5}
\Photon(20,10)(50,30){3}{3}\DashLine(50,30)(80,10){5}
\DashLine(50,70)(50,90){5}
\end{picture}\\
\end{center}

$GG\phi$ vertex:

\begin{center}
\begin{picture}(400,180)(0,0)
\SetOffset(0,90)
\DashLine(20,30)(80,30){5}\DashLine(80,30)(50,70){5}
\DashLine(50,70)(20,30){5}
\Photon(20,30)(20,10){3}{2}\Photon(80,30)(80,10){3}{2}
\DashLine(50,70)(50,90){5}
\SetOffset(80,90)
\DashLine(20,30)(80,30){5}\DashLine(80,30)(50,70){5}
\Photon(50,70)(20,30){3}{5}
\Photon(20,30)(20,10){3}{2}\Photon(80,30)(80,10){3}{2}
\DashLine(50,70)(50,90){5}
\SetOffset(160,90)
\Photon(20,30)(80,30){3}{5}\DashLine(80,30)(50,70){5}
\DashLine(50,70)(20,30){5}
\Photon(20,30)(20,10){3}{2}\Photon(80,30)(80,10){3}{2}
\DashLine(50,70)(50,90){5}
\Text(50,50)[]{f}
\SetOffset(240,90)
\Photon(20,30)(80,30){3}{5}\DashLine(80,30)(50,70){5}
\Photon(50,70)(20,30){3}{5}
\Photon(20,30)(20,10){3}{2}\Photon(80,30)(80,10){3}{2}
\DashLine(50,70)(50,90){5}
\SetOffset(320,90)
\DashLine(20,30)(80,30){5}\Photon(80,30)(50,70){-3}{5}
\Photon(50,70)(20,30){3}{5}
\Photon(20,30)(20,10){3}{2}\Photon(80,30)(80,10){3}{2}
\DashLine(50,70)(50,90){5}
\Text(50,50)[]{f}

\SetOffset(0,0)
\Photon(20,30)(80,30){3}{5}\Photon(80,30)(50,70){-3}{5}
\Photon(50,70)(20,30){3}{5}
\Photon(20,30)(20,10){3}{2}\Photon(80,30)(80,10){3}{2}
\DashLine(50,70)(50,90){5}
\SetOffset(80,0)
\DashLine(20,30)(80,30){1}\DashLine(80,30)(50,70){1}
\DashLine(50,70)(20,30){1}
\Photon(20,30)(20,10){3}{2}\Photon(80,30)(80,10){3}{2}
\DashLine(50,70)(50,90){5}
\SetOffset(160,0)
\DashCArc(50,50)(20,0,360){5}
\Photon(20,10)(50,30){3}{3}\Photon(50,30)(80,10){-3}{3}
\DashLine(50,70)(50,90){5}
\SetOffset(240,0)
\PhotonArc(50,50)(20,0,360){2}{12}
\Photon(20,10)(50,30){3}{3}\Photon(50,30)(80,10){-3}{3}
\DashLine(50,70)(50,90){5}
\SetOffset(320,0)
\PhotonArc(50,50)(20,90,270){2}{6}\DashCArc(50,50)(20,-90,90){5}
\Photon(20,10)(50,30){3}{3}\DashLine(50,30)(80,10){5}
\Photon(50,70)(50,90){3}{2}
\end{picture}\\
\end{center}

$GGG$ vertex:

\begin{center}
\begin{picture}(400,180)(0,0)
\SetOffset(0,90)
\Text(10,100)[1]{}
\DashLine(20,30)(80,30){5}\DashLine(80,30)(50,70){5}
\DashLine(50,70)(20,30){5}
\Photon(20,30)(20,10){3}{2}\Photon(80,30)(80,10){3}{2}
\Photon(50,70)(50,90){3}{2}
\SetOffset(80,90)
\DashLine(20,30)(80,30){5}\DashLine(80,30)(50,70){5}
\Photon(50,70)(20,30){3}{5}
\Photon(20,30)(20,10){3}{2}\Photon(80,30)(80,10){3}{2}
\Photon(50,70)(50,90){3}{2}
\Text(50,50)[]{f}
\SetOffset(160,90)
\Photon(20,30)(80,30){3}{5}\DashLine(80,30)(50,70){5}
\Photon(50,70)(20,30){3}{5}
\Photon(20,30)(20,10){3}{2}\Photon(80,30)(80,10){3}{2}
\Photon(50,70)(50,90){3}{2}
\Text(50,50)[]{f}
\SetOffset(240,90)
\Photon(20,30)(80,30){3}{5}\Photon(80,30)(50,70){-3}{5}
\Photon(50,70)(20,30){3}{5}
\Photon(20,30)(20,10){3}{2}\Photon(80,30)(80,10){3}{2}
\Photon(50,70)(50,90){3}{2}
\SetOffset(320,90)
\DashLine(20,30)(80,30){1}\DashLine(80,30)(50,70){1}
\DashLine(50,70)(20,30){1}
\Photon(20,30)(20,10){3}{2}\Photon(80,30)(80,10){3}{2}
\Photon(50,70)(50,90){3}{2}

\SetOffset(0,0)
\DashCArc(50,50)(20,0,360){5}
\Photon(20,10)(50,30){3}{3}\Photon(50,30)(80,10){-3}{3}
\Photon(50,70)(50,90){3}{2}
\SetOffset(80,0)
\PhotonArc(50,50)(20,0,360){2}{12}
\Photon(20,10)(50,30){3}{3}\Photon(50,30)(80,10){-3}{3}
\Photon(50,70)(50,90){3}{2}
\end{picture}\\
\end{center}

$\phi c\bar{c}$ vertex:

\begin{center}
\begin{picture}(400,90)(0,0)
\DashLine(20,30)(80,30){1}\DashLine(80,30)(50,70){5}
\DashLine(50,70)(20,30){5}
\DashLine(20,30)(20,10){1}\DashLine(80,30)(80,10){1}
\DashLine(50,70)(50,90){5}
\Text(50,50)[]{f}
\SetOffset(80,0)
\DashLine(20,30)(80,30){1}\DashLine(80,30)(50,70){5}
\Photon(50,70)(20,30){3}{5}
\DashLine(20,30)(20,10){1}\DashLine(80,30)(80,10){1}
\DashLine(50,70)(50,90){5}
\SetOffset(160,0)
\DashLine(20,30)(80,30){1}\Photon(80,30)(50,70){-3}{5}
\Photon(50,70)(20,30){3}{5}
\DashLine(20,30)(20,10){1}\DashLine(80,30)(80,10){1}
\DashLine(50,70)(50,90){5}
\Text(50,50)[]{f}
\SetOffset(240,0)
\DashLine(20,30)(80,30){5}\DashLine(80,30)(50,70){1}
\DashLine(50,70)(20,30){1}
\DashLine(20,30)(20,10){1}\DashLine(80,30)(80,10){1}
\DashLine(50,70)(50,90){5}
\Text(50,50)[]{f}
\SetOffset(320,0)
\Photon(20,30)(80,30){3}{6}\DashLine(80,30)(50,70){1}
\DashLine(50,70)(20,30){1}
\DashLine(20,30)(20,10){1}\DashLine(80,30)(80,10){1}
\DashLine(50,70)(50,90){5}
\Text(50,50)[]{f}
\end{picture}\\
\end{center}

$Gc\bar{c}$ vertex:

\begin{center}
\begin{picture}(400,90)(0,0)
\SetOffset(0,0)
\DashLine(20,30)(80,30){1}\DashLine(80,30)(50,70){5}
\DashLine(50,70)(20,30){5}
\DashLine(20,30)(20,10){1}\DashLine(80,30)(80,10){1}
\Photon(50,70)(50,90){3}{2}
\Text(50,50)[]{f}
\SetOffset(80,0)
\DashLine(20,30)(80,30){1}\DashLine(80,30)(50,70){5}
\Photon(50,70)(20,30){3}{5}
\DashLine(20,30)(20,10){1}\DashLine(80,30)(80,10){1}
\Photon(50,70)(50,90){3}{2}
\Text(50,50)[]{f}
\SetOffset(160,0)
\DashLine(20,30)(80,30){1}\Photon(80,30)(50,70){-3}{5}
\Photon(50,70)(20,30){3}{5}
\DashLine(20,30)(20,10){1}\DashLine(80,30)(80,10){1}
\Photon(50,70)(50,90){3}{2}
\SetOffset(240,0)
\DashLine(20,30)(80,30){5}\DashLine(80,30)(50,70){1}
\DashLine(50,70)(20,30){1}
\DashLine(20,30)(20,10){1}\DashLine(80,30)(80,10){1}
\Photon(50,70)(50,90){3}{2}
\Text(50,50)[]{f}
\SetOffset(320,0)
\Photon(20,30)(80,30){3}{6}\DashLine(80,30)(50,70){1}
\DashLine(50,70)(20,30){1}
\DashLine(20,30)(20,10){1}\DashLine(80,30)(80,10){1}
\Photon(50,70)(50,90){3}{2}
\end{picture}\\
\end{center}

\newpage

\end{document}